\documentclass[12pt,preprint]{aastex}
\usepackage{lscape}
\usepackage{verbatim}
\shorttitle{GIs and Solids}
\shortauthors{Boley \& Durisen}

\usepackage{color}
\usepackage{ulem}

\begin{document}
\title{On The Possibility of Enrichment and Differentiation in Gas Giants During Birth by Disk Instability}

\author{Aaron C.\ Boley \& Richard H.~Durisen}

\affil{Department of Astronomy; University of Florida,
211 Bryant Space Science Center, Gainesville,
FL 32611, USA; aaron.boley@gmail.com}
\affil{Department of Astronomy, Indiana University,
727 East 3rd Street, Swain West 319, Bloomington, IN 47405, USA}

\begin{abstract}


We investigate the coupling between rock-size solids and gas during the formation of gas giant planets by disk fragmentation in the outer regions of massive disks.
In this study, we use three-dimensional radiative hydrodynamics simulations and model solids as a spatial distribution of particles.   We assume that half of the total solid fraction is in small grains and half in large solids.  The former are perfectly entrained with the gas and set the 
opacity in the disk, while the latter are allowed to respond to gas drag forces, with the back reaction on the gas taken into account. To explore the maximum effects of gas-solid interactions, we first consider 10cm-size particles.  We then compare these results to a simulation with 1 km-size particles, which explores the low-drag regime. 
We show that (1) disk instability planets have the potential to form large cores due to aerodynamic capturing of rock-size solids in spiral arms before fragmentation; (2) that temporary clumps can concentrate tens of $M_{\oplus}$ of solids in very localized regions before clump disruption; (3) that the formation of permanent clumps, even in the outer disk, is dependent on the grain-size distribution, i.e., the opacity; (4) that nonaxisymmetric structure in the disk can create disk regions that have a solids-to-gas ratio greater than unity; (5) that the solid distribution may affect the fragmentation process; (6) that proto-gas giants and proto-brown dwarfs can start as differentiated objects prior to the H$_2$ collapse phase;  (7) that spiral arms in a gravitationally unstable disk are able to stop the inward drift of rock-size solids, even redistributing them to larger radii; and, (8) that large solids can form spiral arms that are offset from the gaseous spiral arms.  We conclude that planet embryo formation can be strongly affected by the growth of solids during the earliest stages of disk accretion. 

\end{abstract}

\section{Introduction}
Core accretion (e.g., Pollack et al.~1996; Hubickyj et al.~2005) and direct formation by disk instability (e.g, Cameron 1978; Boss 1997, 1998) are both viable modes of gas giant planet formation, and both may be required to explain the diversity of observed planetary systems.    For disk instability to lead to fragmentation, the Toomre (1964) $Q$ parameter must be rapidly lowered in the outer disk, which is where the disk is most likely to have both a low $Q$ and short cooling times relative to the local orbital time (e.g., Rafikov 2007, 2009; Nero \& Bjorkman 2009; see Durisen et al.~2007 for a review). Although disk instability is quite sensitive to the details of radiative transfer approximations and to opacity (e.g., Johnson \& Gammie 2003; Cai et al.~2006, 2008, 2009; Boley et al.~2007b;  Mayer et al.~2007; Cossins et al.~2009), hydrodynamics simulations of extended, outer disks demonstrate that fragmentation is possible even under nonisothermal conditions (e.g., Voroybov \& Basu 2006, 2010; Stamatellos et al.~2007; Stamatellos \& Whitworth.~2009; Stamatellos et al.~2009; Boley 2009; Forgan et al.~2009; Hayfield  et al.~2010).   If fragmentation is to occur, it is likely to happen during the earliest stages of disk evolution, when mass infall onto the disk can proceed faster than slow, diffusion-like mass transport  (e.g., Boley 2009; hereafter B2009).  This mass loading will likely lead to the development of global spiral modes, nonlocal mass transport, and in some cases, fragmentation.   
This led B2009 to suggest two modes of gas giant planet formation, with disk instability representing the early mode, operating in newly-formed systems, and with core accretion representing the dominant mode during the evolved stages of disk evolution.  An example of a system with disk instability planets may very well be HR8799 (Marois et al.~2008), as other formation scenarios seem to be unlikely (Dodson-Robinson et al.~2009).

Several observational signatures have been proposed to test whether disk instability plays a role in planet formation. B2009 suggested that a bimodal population of gas giants would reflect that both formation channels are possible and that the ratio of wide orbit gas giants to the total number of planets in a system should increase with decreasing metallicity.  The latter expectation reflects that core accretion should be more sensitive to changes in metallicity than disk instability.  If, instead, wide orbit gas giants are produced mainly by scattering, Dodson-Robinson et al.~(2009) pointed out that  the frequency of systems with gas giants on wide orbits should be dependent on system age.  As more systems with wide orbit gas giants are discovered, these predictions will be tested against observations.

In addition to population trends, metallicity and core size may be a way to distinguish between formation mechanisms. For example,  Helled \& Bodenheimer (2010) use a planetary evolution code to calculate contraction times for isolated clumps, and follow planetesimal capture of large bodies ($\ge1$ km) until molecular hydrogen dissociates in the clump's core. This dissociation causes the clump to collapse rapidly to about a Juptier radius, which significantly reduces the cross section of the young planet and marginalizes planetesimal capture.  They find that planetesimal capture can only contribute negligible to modest enhancement of solids for planets at large radii, concluding that planets that form on wide orbits by disk instability should have nearly the same composition as their host star.  If this is correct, metallicity could be a valuable tool in determining formation modes for wide orbit gas giants, as gas giant planets that form by core accretion are expected to be metal rich (e.g., Pollack et al.~1996).  However, a limitation of the Helled \& Bodenheimer calculation is that their simulations are not within the context of a gravitationally unstable disk.  When clumps form, they tend to form near corotation or at intersections of spiral arms (Durisen et al.~2008), yielding initial fragment masses that contain material from a small segment of the spiral arm (Boley et al.~2010).

Because  spiral arms are the most likely regions for fragmentation, one must also consider enrichment of solids at birth.  Solids are expected to move relative to the gas due to the lack of pressure support (Weidenschilling 1977).  Depending on the direction of the pressure gradient, this disparity between gas and particle forces can lead to either  a tail or headwind for a given particle. The wind exchanges angular momentum with the solid and causes it to migrate.   Haghighipour \& Boss (2003) showed that this effect will cause particles to become trapped in local pressure maxima, and Rice et al.~(2004, 2006) used SPH calculations with solid particles to demonstrate that spiral arms in a gravitationally unstable disk are ideal places for trapping solids.  These results suggest that enrichment at birth could provide newly-formed fragments with a supernebular solids-to-gas ratio.  In addition,  such concentration of solids could provide enough material to form substantial rocky/icy cores, regardless of whether the fragment survives in the disk.

In this paper, we present hydrodynamics+particle simulations to explore the heavy-element enrichment  of fragments at birth in spatially extended, massive disks during the accretion phase. In Section 2, we outline the numerical methods for this study, including introducing our algorithm for evolving solids.  We then describe the numerical experiments and disk initial conditions in Section 3.  We present our results in Sections 4 and 5, and discuss the implications of some of these results in section 6.  We summarize our conclusions with a bulleted list in Section 7.

\section{Method}

\subsection{Hydrodynamics}

We use CHYMERA (Boley 2007), with the particle modifications described below, to run a series of hydrodynamics simulations of gravitationally unstable disks with a significant population of rocks.  CHYMERA is an Eulerian code that solves the equations of hydrodynamics, with self-gravity, on a regular, cylindrical grid.  The star's motion is solved self-consistently, as described in B2009.  The internal energy of the gas accounts for the vibrorotational states of H$_2$ as described in Boley et al.~(2007a).  Because we are studying extended disks that have very long orbital timescales, we assume that the ortho-para ratio of H$_2$ remains in equilibrium for the local temperature.  For radiative cooling, CHYMERA can use flux-limited diffusion (Boley et al.~2006), a hybrid ray-tracing+flux-limited diffusion scheme (Boley et al.~2007b), or a local radiative cooling approximation (B2009).  The hybrid scheme is the most accurate, but because it solves for radiative transport explicitly, it can become unstable whenever the radiative time step becomes smaller than the hydrodynamics step.  Under most conditions, the algorithm is stabilized by limiting the amount of energy a cell can change for a given step.  This sacrifices accuracy for speed, but appears to be reasonable for many situations (see Boley 2007).  To avoid using limiters, a subcycling version of the hybrid scheme has been developed, but it is still in its testing phase and was not ready for use when this study began.   Because we are interested in how the solids distribution behaves in a disk that does fragment, we choose to use the fast and stable local radiative cooling approximation.    

In order to suppress artificial fragmentation, we ensure that the Truelove et al.~(1997) criterion (see also Nelson 2006)  is obeyed at all times by adding a cold pressure floor to our equation of state, i.e., $P=\max(P_{\rm ideal},P_{\rm cold})$. We have chosen to set the minimum Jeans wavelength to be at least four radial grid cells, which sets $P_{\rm cold}= (4 \Delta r)^2 G \rho^2/\pi$.  As will be discussed in more detail below, the cold pressure only becomes necessary at the very center of clumps.  

\subsection{Gas Drag Algorithm}

For modeling solids, we have augmented CHYMERA to include particles, where each particle represents an ensemble of solids with some radius $s$.  The Epstein and Stokes drags are calculated separately, and an interpolated solution is then found.  First consider the Epstein limit, which is appropriate whenever the Knudsen number ${\rm Kn}=\lambda/2s\gg1$.  Here,  $\lambda= \mu m_p (\pi R^2 \rho)^{-1}$ is the mean-free-path of a gas particle, where $\rho$ is the gas density with mean molecular weight $\mu$, $m_p$ is the proton mass, and $R$ is the typical gas atom/molecule radius.  The internuclear spacing of H$_2$ is $r_0\sim 0.74 \AA$, and the typical diameter for a molecule is $\sim 3r_0$ (Allen 1976), so we take $R=10^{-8}$ cm.  This size-scale is slightly smaller than what has been used in other work (cf.~Rice et al.~2004), but gives a cross section that is within a factor of a few to expected cross sections for atomic and molecular gas ($\sim 10^{-15}$ cm$^{2}$; Allen 1976).   

We set the velocity difference between the gas and solids to be $\delta v=v_g-v_p$. Following Paardekooper (2007), we write the change of this difference in the Epstein limit as
\begin{equation}
\frac{d \delta v}{dt}\bigg\vert_{\rm Epstein}=-\left(\frac{8}{\pi}\right)^{1/2}\frac{\rho c_a}{\rho_s s} f_e\delta v,
\end{equation}
for adiabatic sound speed $c_a$ and internal particle density $\rho_s$. 
The term $f_e$ is used to interpolate smoothly between low and high-Mach speed flows, where the Mach speed here is defined relative to the gas such that $\mathcal{M}=\delta v/c_a$ and $f_e=(1+9\pi\delta v^2/(128 c_a^2))^{1/2}$ (Kwok 1975).  
It is also convenient to define the stopping time for a particle moving through the gas as (e.g., Weidenschilling 1977)
\begin{equation}
t_s=\frac{\rho_s s}{\rho c_a}.
\end{equation}
Assuming that the gas density and sound speed are constant over a time step $\Delta t$,  equation (1) can be integrated analytically. Carrying out the integration, we find
\begin{equation}
\delta v\bigg\vert_{\rm Epstein} = 2 \delta v_0 \beta \mathcal{Q} \exp\left(-\alpha \Delta t\right) \left(\mathcal{Q}^2-\delta v_0^2\exp\left(-2\alpha \Delta t\right)\right)^{-1},
\end{equation}
where $\delta v_0$ is the initial velocity difference, $\alpha=\left( \frac{8}{\pi}\right )^{1/2} \frac{\rho c_a}{\rho_s s}$,  $\beta=\left(128c_a^2/(9\pi)\right)^{1/2}$, and $\mathcal{Q}=\beta+\left(\beta^2+\delta v_0^2\right)^{1/2}$ 
(see also Paardekooper \& Mellema 2006 for an equivalent expression).

Now consider the Stokes limit, which is appropriate for small Kn.  The change in the velocity difference can be written as
\begin{equation}
\frac{d\delta v}{dt}\bigg\vert_{\rm Stokes}=-3\alpha {\rm Kn} k_d \delta v.
\end{equation}
The factor $k_d$ depends on the Reynolds number ${\rm Re}=3\left(\frac{\pi}{8}\right)^{1/2}\frac{\mathcal{M}}{\rm Kn}$, where (e.g., Paardekooper 2007)
\begin{equation}
k_d=\left\{ \begin{array}{l@{\quad  \quad}l}
1+0.15 {\rm Re}^{0.687}  & {\rm for~Re\le500,} \\
3.96\times10^{-6}{\rm Re}^{2.4} &{\rm for~Re \le 1500,}\\
0.11{\rm Re} &{\rm otherwise.}\end{array}\right. 
\end{equation}
Each Reynolds regime can be integrated separately, yielding the following solutions:
\begin{equation}
\delta v \bigg\vert_{\rm Stokes}=\left\{ \begin{array}{l@{\quad  \quad}l}
\delta v_0 e^{ -3\alpha{\rm Kn} \Delta t}\left( \Theta \vert \delta v_0 \vert ^{0.687}\left(1-e^{-2.061 \alpha {\rm Kn}\Delta t}\right)+1\right)^{-1.4556} & {\rm for~Re\le500,} \\
  \delta v_0 \left(1+7.2 \vert \delta v_0 \vert^{2.4} {\rm Kn} \alpha \Upsilon \Delta t  \right)^{-0.4167} &{\rm for~Re \le 1500,}\\
\delta v_0 \left(1+0.99 \left(\frac{\pi}{8}\right)^{1/2} \frac{\vert \delta v_0 \vert\alpha}{c_a}\Delta t\right)^{-1}  &{\rm otherwise.}\end{array}\right. 
\end{equation}
For convenience, we have defined $\Theta = 0.15 \left(\left(\frac{\pi}{8}\right)^{1/2}\frac{3}{c_a \rm Kn}\right)^{0.687}$ and $\Upsilon=3.96\times10^{-6}\left(\left(\frac{\pi}{8}\right)^{1/2}\frac{3}{c_a \rm Kn}\right)^{2.4}$, which are constant over a time step.

Finally, the Epstein and Stokes solutions can be coupled to find an interpolated solution  (modified from Woitke \& Helling 2003)\footnote{The interpolation weights used by Woitke \& Helling are $\frac{(3{\rm Kn})^2}{(3{\rm Kn}+1)^2}$ and $\frac{1}{(3 {\rm Kn} +1 )^2}$.  However, they apply these weights directly to the force, where we solve for the velocity difference for each limit independently and seek a coupled solution.  For this reason, we modify the weights  to sum to unity. The discrepancy in the sum of the weights is, at worst, different by a factor of two at ${\rm Kn=1/3}$.  Figure 21 in the Appendix shows that drift velocities are captured with accuracy sufficient enough for this study.}
\begin{equation}
\delta v \bigg\vert_{\rm total}=\frac{\left(3{\rm Kn}\right)^2}{\left(3{\rm Kn}\right)^2+1} \delta v\bigg\vert_{\rm Epstein}+\frac{1}{\left(3{\rm Kn}\right)^2+1}\delta v\bigg\vert_{\rm Stokes}.
\end{equation}
Using conservation of momentum, this new velocity difference can be used to find the new momentum and velocity for the gas and solids.  We do include the back reaction of the particles on the gas in this algorithm.

We should point out that a limitation of our approach is that the analytic 
solutions for the velocity difference consider only the drag force.  However, differences between forces on the gas compared with those on the particles due to pressure require that either these forces be included in the solution of the velocity difference or that the hydrodynamics time step remains smaller than the stopping time of the particles.  If these requirements are not met, then the asymptotic velocity drift of the particles through the gas will be modeled poorly.  For the particle sizes and density regimes we consider here, this should not be a major difficulty.  We present a drift test in the Appendix that addresses this issue. 

\subsection{Coupling Gas and Solids}

We use a crude particle-in-cell method for evolving the particles.  At the start of the simulation, particles are assigned to cells randomly, weighted by the gas mass in a given cell.  Inside the cell, each particle is given a random position, an orbital frequency that is Keplerian, and a random vertical and radial velocity as explained in Section 3.2. 
Whenever the density of solids is needed, it is calculated by summing the mass of all particles in a given cell and dividing by that cell's volume.  As discussed below, this density is included when calculating the disk's self-gravity.  For future studies, we would like to include density estimates using a smoothing kernel over nearest neighbors, but that is not implemented here. 

We integrate the particles using a leap-frog method in Cartesian coordinates. During a given  hydrodynamic
step $\Delta t$, the particles are integrated after the second hydrodynamics sourcing step, which occurs after the potential update (see Figure 2.2 in Boley 2007), using the following scheme:
\begin{eqnarray}
v_0&=&v_{-1/2}+\frac{1}{2}a_{\rm grav} \Delta  t_{-1} \\ \nonumber
v_{0}'&=&G_{\rm drag}(\delta v, v_0; \Delta t_{-1}) \\ \nonumber
v_{1/2}&=&v_{0}'+\frac{1}{2}a_{\rm grav} \Delta t_{0} \\ \nonumber
v_{1/2}'&=&G_{\rm drag}(\delta v, v_{1/2}; \Delta t_{0})\\ \nonumber
x_{1}&=&x_0+v_{1/2}'\Delta t_0.
\end{eqnarray}
Here, $v$ and $x$ are the velocity and position values in Cartesian coordinates.  The subscript 0 represents the values at the beginning of the current step.  Using the previous time step with the current acceleration for the first update allows for better accuracy during variable time stepping.  The velocity is first updated using the acceleration due to gravity.  Next, using the updated velocity, a new velocity $v'$ is found by applying the drag force using equations (3), (6), and (7) and using conservation of momentum (see below), all represented in equation (8) as $G_{\rm drag}$.    After the particle velocities are updated, the positions  are advanced using $v'$.  This algorithm permits $10^5$
particles to be evolved with a modest performance penalty.

For calculating $a_{\rm grav}$, we first apply the exact force on the particle due to the star using Cartesian coordinates. 
Next, the force due to self-gravity from the disk is calculated on the cell faces of the cylindrical grid.  These forces are then interpolated linearly, in the same way as the velocities (see below), to the particle's position in cylindrical coordinates.  These interpolated values are then transformed to Cartesian coordinates and added to the star's force.   
The drag force is calculated by transforming the $i$th particle's Cartesian velocities to cylindrical components $v_{r,i}$, $v_{z,i}$, and $v_{\phi,i}$.  The initial velocity differences used in equations (3), (6), and (7) are then calculated by linearly interpolating the gas velocities to the position of the particle. These interpolations, for now, are only one dimensional.  For example, regardless of a particle's $\phi$ or $z$ position within a given cell, the face-centered radial velocity component is used for interpolating to the $r$ position for finding the gas $v_{r}$.  Likewise, regardless of a particle's $\phi$ or $r$ position in a cell, the face-centered vertical velocity component is used for interpolating to the $z$ position for finding the gas $v_z$.  When calculating the local gas orbital frequency $\Omega$ within the $J$th radial cell, we use $\Omega=\Omega_J+(\Omega_{J+1}-\Omega_{J-1})(r_i-r_J)/(2 \Delta r)$, regardless of the particle's $\phi$ or $z$ position within the given cell, for particle 
radial position $r_i$ and cell center position $r_J$.  This interpolation scheme is coarse but sufficient for this study (see Appendix). Using the time step $\Delta t/2$, the new $\delta v$ is calculated for each component.  So far, we only have the new velocity differences.  We now use conservation of momentum to find the new velocity for the particle and for the gas, as affected by this particle.  For example, in the radial direction, we have 
$\delta v_{r,0} = v_{r,g,0}-v_{r,p,i,0}$ and initial 
total radial momentum density
$S=\rho_{g}v_{r,g,0}+\rho_{p,i} v_{r,p,i,0}$,
where $v_{r,g,0}$ is the current gas radial velocity at the interpolated position, $\rho_g$ is the gas density for the cell (cell center value), and $v_{r,p,i,0}$ is the $i$th particle's radial velocity. The density for the $i$th particle, $\rho_{p,i}$, is found by dividing the particle's mass by the volume of the cell that contains the particle. The naughts represent values before the drag from the $i$th particle is applied. 
The updated $\delta v_r$ is then found by using $\delta v_{r,0}$ in equations (3), (6), and (7).  The new velocity in the radial direction for the $i$th particle is then $v_{r,p,i}=(S-\rho_g\delta v_r)/(\rho_g+\rho_{p,i})$.
The new gas momentum at the location of the particle is then calculated.  In the radial and vertical directions, the momentum densities are updated on cell faces by extrapolating the 
change in the gas momentum density to the cell faces using the same interpolation weights that were used to calculate the gas velocity at the particle's position.  This is done to maintain the staggered grid formalism used in the code. Because the angular momentum density is cell centered, the change in the gas's angular momentum is applied directly to the cell in which the particle is contained. 
Finally, we should make clear that every time the drag force is applied, the velocities for both the gas and the $i$th particle are updated.  If the $i$th$+1$ particle resides within the same cell as the $i$th particle, then the drag on the $i$th$+1$ particle will be affected by the drag update for the $i$th particle.  Waiting to update the gas velocities until all particles have had their drag forces taken into account leads to numerical instabilities when the particle density approaches the gas density.

\section{Initial Conditions}

\subsection{Gas Model}

A principal objective of this study is to address whether clumps that form by disk instability can be enriched at birth.  To proceed, we explore two disks, each of which is forced to become highly unstable in order to effect fragmentation in the system.   The first initial condition (IC1) is a 0.4 $M_{\odot}$  disk around a 1.5 $M_{\odot}$ star.  We use a softening for the star of  $5 AU$  as described in B2009.  The ICs were created using the analytic approximations described in Boley \& Durisen (2008).  Unfortunately, this approximation uses Keplerian rotation, assumes a polytropic vertical temperature profile, and has sharp inner and outer boundaries.  As a result, the analytic disk was evolved at low azimuthal resolution for about two orbital periods at $r\sim200$ AU to allow the disk to adjust.  During this evolution, the disk was irradiated using a temperature profile of $T_{\rm irr}=T_0 (r /\rm{AU})^{-1/2} + 10$ K, with $T_0=600$ K.   Figure 1 shows the corresponding Toomre (1964) $Q$ and surface density $\Sigma$ profiles.  Recall that a disk is unstable to gravitational instabilities whenever $Q=c_a\kappa/(\pi G\Sigma)\lesssim 1.7$, where $\kappa$ is the epicyclic frequency and $c_a$ is the adiabatic sound speed (e.g., Durisen et al.~2007).   In general, instability does not mean fragmentation, as GIs are capable of reaching a state of self-regulation (e.g., Gammie 2001; Lodato \& Rice~2004; Mejia et al.~2005), even when radiation transport is included (Nelson et al.~2000; Boley et al.~2006, 2007;  Stamatellos \& Whitworth 2008).  However, at least for isothermal disks, fragmentation becomes very likely whenever $Q<1.4$ (e.g, Nelson et al.~1998; Mayer et al.~2004). Figure 1 shows that  the high irradiation temperature used for the initial model stabilizes the disk against fragmentation, but not against  the development of gravitational instabilities (GIs).  During the low-resolution evolution, the surface density profile has relaxed to an exponential that roughly follows $\Sigma\sim 410 \exp(-r/45 {\rm AU})$.  This is not a detailed fit, as we are looking at a small range of disk radii, but it does give the basic sense of the surface density profile. Hereafter, simulations that use IC1 are labeled SIM1. 

At the start of the simulation, SIM1's ICs are loaded onto a high-resolution grid, with $r$, $z$, $\phi$ = 256, 64, 512 zones.  The initial radial and vertical cell widths remain unchanged from the relaxation process, with $\Delta r=\Delta z=2$ AU. The disk is given a 10\% random density perturbation, and the prefactor $T_0$ in the irradiation temperature profile is reduced to 130 K.    Dropping the irradiation temperature by such a large factor causes a sudden loss of disk stabilization and allows $Q$ to drop below unity temporarily.  The decision to change $T_0$ dramatically is mainly to drive SIM1 to instability violently; however, it does have the physical interpretation of sudden and prolonged shielding of the outer disk by, for example, the inner disk.

The second set of ICs (IC2) uses the B2009 simulation called SIMD.  We start our realization of the simulation about half of an orbit before fragmentation (Fig.~2).  To recapitulate, the disk is 0.33 $M_{\odot}$ and surrounds a 1 $M_{\odot}$ star.  We do not use softening for the interactions between the star and gas in this simulation.  The grid has the same dimensions and physical resolution as the SIM1 grid.  The disk is grown by allowing mass accretion from an envisaged envelope at $\dot{M}\sim10^{-4} M_{\odot}$ yr$^{-1}$ until 0.33 $M_{\odot}$ is reached, after which the accretion is abruptly halted. 
This mass accretion rate is high, and is chosen mainly as 
a computational time consideration (see B2009 for more details). The radial density profile for the infalling mass is set to $\rho\sim r^{-0.5}$ (note that $r^{-1.5}$ profiles were also explored in B2009).  This shallow density profile causes mass to build up at large radii, which creates a  surface density enhancement in the disk's profile (Fig.~3) and causes $Q$ to plummet. The surface density profile reflects that, for this particular case, disk formation occurs faster than GIs can activate and smooth out the density structure.  During the formation of the disk and its subsequent evolution, an irradiation temperature of 30 K is incident on the disk everywhere. We include this irradiation for the simulation presented here, which we call SIM2.

\subsection{Solids and Opacity Assumptions}

We assume that half of the solid mass is contained in large particles that do not contribute to the opacity and that  half goes into a perfectly gas-entrained dust distribution that is solely responsible for the opacity. We use the D'Alessio et al.~(2001) opacity tables, and only consider the Rosseland mean for simplicity.  The grain size distribution is assumed to be a power law, with $dn/ds\propto s^{-3.5}$ and a minimum grain size of $0.005\micron$.  
The maximum grain size in the distribution can be varied, which can have a strong effect on the cooling (e.g., Cai et al.~2006).  SIM1 is run with a maximum grain size of 1 mm and  with 1 $\mu$m, while SIM2 is only run with 1 $\mu$m. Hereafter, 
when we use the term ``1 mm opacities,''  we are referring to a dust grain distribution with a maximum $s=1$ mm.  The total  solids-to-gas 
mass ratio is assumed to be 0.02 (including ices), with 0.01 in small grains, which sets the opacity, and 0.01 in rock-size solids composed of silicates and ices. We refer to the solids in this manuscript simply as ``rocks'' because most simulations are run with 10cm-size particles, but rocks can also refer to the solids in the 1km-size run.    ÊWe choose this simple division between  large solids and grains
for multiple reasons, which are detailed at the end of this subsection.  Note that SIMD in B2009 was assumed to have solar metallicity, so our assumption here reduces the original dust opacity for SIM2 by a factor of two.  

Particles are distributed according to a density-weighted Monte Carlo sampling of the disk.  The initial particle surface densities are plotted in Figures 1 and 3 for SIM1 and SIM2, respectively, and show good agreement with the gas surface density.  We assume that each particle in the simulation represents an ensemble of solids. The particles in any given run are assumed to have a single size population of either $10$ cm or 1 km in radius and an internal density of 3 g cm$^{-3}$.   For the initial velocity distribution, the particles are placed on Keplerian orbits for the star.  The disk potential in the initial orbital frequency was not included because we intended to focus on just 10-cm size particles, which respond strongly to gas drag.  The 1 km-size run was originally intended to be presented in a complementary follow-up study, but for providing a comparison with the low-drag limit, it was later decided to run SIM2 with km-size particles. To make sure that any difference we find can be attributed to the change in solid size only, we also put the 1 km-size particles on Keplerian orbits.  In SIM1, the radial and vertical velocities were set to zero and given a random perturbation with a magnitude no greater than 0.01 of the Keplerian speed.  The perturbation was included in an attempt to avoid all particles at a given disk radius from passing through the midplane simultaneously. In SIM1, we found the 0.01 perturbation to be too weak, so the perturbation was increased to 0.1 of the Keplerian speed for SIM2. The larger perturbation kept strong caustics from forming during particle midplane passages. However, the fast midplane settling times ($\sim 1$ orbit) marginalize this detail for the 10 cm-size solids.  

For this study, we do not change the fraction of solids that go into dust grains.  Although our 50-50 choice is arbitrary, 
the effects of decreasing or increasing the amount of dust in the system on GI amplitudes have already been addressed in previous research (see Cai et al. 2006, 2008; Boley \& Durisen 2008).  As more solids remain in a dusty phase, the opacity of the disk will increase, which will tend to stabilize the disk against fragmentation.  Likewise, as the opacity is lowered due to more dust going into larger solids, the amplitude of GIs can increase due to shorter cooling times, as long as the emissivity does not approach zero. Another consideration is whether there is a qualitative change in the disk's behavior for different large solids-to-gas ratios, which could occur for different disk metallicities or for different grains/large solids fractions. The 50-50 choice seems to be a reasonable starting point, as it is not at either extreme, and, as we will show, does demonstrate enrichment and differentiation at birth. To begin to understand how the large solids affect the large-scale structure of the gaseous disk, we run one simulation without large solids, while keeping the opacity the same.

We explore two particle sizes, but not a distribution of particle sizes.  There are several reasons for this choice.  First, we seek to run as few supercomputer simulations as necessary to demonstrate the enrichment of fragments at birth. We feel this can be done well by picking a solid size that is likely to maximize enrichment and differentiation at birth, which is represented by 10 cm rocks  (see the appendix).  Particles that are much smaller will remain entrained with the gas, while particles much larger will be unaffected by gas drag. The former does not require additional simulations, as this has been the assumption in most GI hydrodynamics simulations to date. The latter is demonstrated using the 1 km solid size simulation, but it is not expected to lead to significant enrichment (e.g., Helled \& Bodenheimer 2010).  Instead of exploring a single population, one might argue that a distribution of particle sizes would be more realistic.  While this is correct qualitatively, any distribution that we could have chosen would be as equally arbitrary as a single solid size inasmuch as the
actual distribution of solid sizes is unknown, particularly during the early stages of disk evolution explored here. Even during later stages, planetesimal growth is a matter of ongoing research (e.g.,
Johansen et al.~2007; Morbidelli et al.~2009).  Consider, for example, the results of the Brauer et al.~(2008) study.  They find that the particle size of solids can become strongly peaked between 1 and 10 cm at 100 AU for the conditions they explore, similar to what is explored in this manuscript, but only if they ignore collisional destruction as well as inward drift due to gas drag.  If they include destruction and drift, they find very little dust growth, which is inconsistent with observations.  For example, Greaves et al.~(2008) and Ricci et al.~(2010) both find evidence for particles with sizes of a few cm or larger in the extended regions of disks.  Even if we use these observations to infer a power law distribution for particle sizes up to 10 cm (e.g., Draine 2006), we are still are left without knowing the distribution of solids larger than 10 cm, due to observational limits, or whether the distribution appropriate for disks with ages of a few Myr is also appropriate for very young disks. Finally, there is a danger in assuming some distribution of solids without having a model for grain growth and destruction that can be evolved self-consistently.  As solid sizes differentiate due to disparate drag timescales, the local population of solids will also evolve.  The link between the spatial and size evolution of the solids will alter the differentiation process itself, as well as local opacities.  This problem is by no means removed by considering a single particle size, as a local concentration of rocks will likely change the local dust opacity, but a single size does allow as clean of a numerical experiment as possible.  Complexity can be added in later studies after the simple case is understood, which we present in detail here.

\subsection{Simulation Nomenclature}

Overall, four variations of the simulations are run.  As is described in the next section, the SIM1 disk is evolved with 1mm opacities, with 1$\micron$ opacities, and 
the same $1\micron$ opacities, but without rocks.  We refer to these simulations as SIM1mm, SIM1mu, and SIM1muNoRocks, respectively.  SIM1mu and SIM1muNoRocks are both branched from the SIM1mm disk after the first 620 yr, as described below.  When referring generally to characteristics of the disk, we only write SIM.  Two realizations of SIM2 are presented in this study, one using 10 cm-size particles (simply SIM2) and 1 km-size particles (SIM2km), both of which are run with 1$\micron$ opacities.

\section{Results}

\subsection{SIM1: Gas Evolution}

Due to the short cooling times, the sudden change from $T_0=600$ K to 130 K causes the disk to contract quickly, and after about 620 yr of evolution, the SIM1 disk has contracted enough that it becomes possible to evolve the same disk on a grid with half the initial physical size in $r$ and $z$.  Using direct injection, i.e., no interpolation, the disk is loaded onto a higher resolution grid with $\Delta r=\Delta z=1$ AU, while keeping the number of radial, azimuthal, and vertical grid cells the same.  Figure 4 shows the $Q$ and surface density profiles shortly after being loaded onto the high physical resolution grid.  The surface density profile has been altered significantly after reducing the irradiation temperature.  The choppy $Q$ profile is due to direct injection, which adds noise to the epicyclic frequency.  According to Figure 4, the disk is strongly unstable.


SIM1mm is only run for 1260 yr.  The surface density evolution is shown in Figure 3.  Although very flocculent structure develops, including knot formation, clumps do not form. If, however, the maximum grain size is reduced to 1 $\mu$m at 620 yr, which marks the start of SIM1mu, the disk fragments into many clumps.  At the time of the opacity switch, dense spiral arms have not yet fully developed. Figure 6 shows the evolution of SIM1mu from 620 yr to 1870 yr.  The effect of opacity on the spiral structure is illustrated by the $\rho$-$T$ phase diagram in Figure 7.  For the same temperature, SIM1mu reaches much higher densities than SIM1mm.   Figure 8 shows the D'Alessio Rosseland mean opacity for a maximum grain size of 1 $\micron$ and 1 mm.  For $T\lesssim 100$ K, the 1 mm opacity is considerably more opaque, which does not allow the spiral arms in SIM1mm to cool fast enough for fragmentation to proceed.  One might worry, instead, that the cold pressure (see section 2.1) employed in these simulations could be affecting fragmentation.  The thick line in Figure 7 shows where the cold pressure becomes important, assuming an equivalent thermodynamic temperature for the pressure.  This threshold is only met in clump cores, so this cold pressure cannot be responsible for the lack of fragmentation in the 1 mm case.   It is possible that, after significantly longer evolution, SIM1mm 
would fragment, too.  We chose not to investigate this possibility for two reasons: (1) We wanted to focus our computing resources on disks that we know will fragment; and
(2) even this short simulation demonstrates that the grain size in the outer disk can strongly affect clump formation.

Finally, before focusing on the evolution of the rocks in these simulations, we compare disk structures between SIM1mu and 
SIM1muNoRocks (Fig.~9).    Although the disks are qualitatively similar, differences have developed. This will be discussed in more detail in the next section.

\subsection{SIM1: Rock Evolution}

By the time spiral arms in the gas form, most rocks have settled into a very thin subdisk, contained within one vertical grid cell.  This is consistent with expectations: As discussed in the Appendix, the stopping time for 10 cm rocks is $\sim 10$-$100$ yr for the gas densities in these disks.  Most particles will also make their first midplane passage after 1/4 of an orbit due to the ICs, i.e.,  about 200 yr at 100 AU.  As seen in Figure 6, the spiral arms remain weak until about 780 yr, giving enough time for particles to settle before spiral waves fully develop.

Figure 10 shows the face-on evolution of the gas and particles in SIM1mu.  As rocks encounter gas over-densities in spiral arms, they become trapped and concentrated, which is consistent with the results reported by Rice et al.~(2004).  By the end of the simulation, most rocks are associated with spiral arms or clumps. To determine whether there is an enhancement of solids in these structures, we plot in Figure 11 the ratio of rock surface density to gas surface density for four snapshots.  The data are shown using the cylindrical coordinates of the grid, where the abscissa is the $r$ coordinate and the ordinate is the $\phi$ coordinate, to marginalize the chance that a high peak in the ratio is due to interpolation error.  In about one dynamical time, the solids become filtered and caught in the spiral arms, with some regions exceeding a tenth of the gas surface density.   The right panel in Figure 11 shows the rock-to-gas ratio for  gas and particles in only the midplane cells, which accounts for most of the rock mass.   Thin, wispy rock arms dominate the midplane density in some regions.

 In addition to concentrating the rocks into arms, the gas-rock coupling in this highly perturbed disk stops rapid inward drift.  Rocks still move inward, but they are locked with the progression of spiral arms.  Figure 12 (left panel) shows the azimuthally averaged surface density evolution for the rocks.  For the first $\sim 600$ yr of evolution, particles outside $r\sim 70$ AU move inward relative to the gas, while particles inside 70 AU move outward and collect in an annular surface density maximum.  After $\sim 600$ yr, spiral arms develop in the 70 AU surface density enhancement, and undermine simple inward drift  due to nontrivial gas dynamics and the evolving pressure maxima.   Note that rocks move inward and outward over tens of AU.  The spikes in the rock surface density correspond to gaseous spiral arms and/or clumps.  As shown in Figure 12 (right panel), the surface density enhancements in rocks coincide with maxima in the azimuthally averaged gas surface density.  
 
 Occasionally, features that do not have an obvious gas surface density counterpart form.  The most obvious type of feature is a rock bridge between gaseous spiral arms.  These structures are created when a gas spiral arm dissipates, but the thin, dense rock arm does not.  As the spiral arm dissipates, it does so by diffusing away from the pressure maximum.  This would allow for particles that are highly concentrated and not perfectly entrained with the gas to remain concentrated.   Bridges mostly have irregular shapes, but rare, straight bridges do form after interacting arms create a temporary leading rock arm that is then sheared into a trailing arm.  Knotty structure in the rocks can also form due to buckling of the rock arm after incomplete dissolution of the gaseous arm (see $x,y\approx 50,50$ AU in the 1410 yr panel of Fig.~10). 
 
 Finally, the strong clumping and density enhancements in the rock distribution seem to be causing noticeable changes in the evolution of the gas.  In particular, as seen in Figure 9, SIM1mu shows additional clumping and fragmentation compared with SIM1muNoRocks.   Unfortunately, it is difficult to distinguish whether the differences between the simulations are brought about by the chaotic nature of GIs in the nonlinear regime, or whether they are directly related to the solids.  This is particularly difficult because variations of a percent could be amplified over a dynamical time for either reason.  For example, by removing the solids from the simulation, 
we instantaneously decrease the mass of the disk by one percent.  On the other hand, significant concentration in spiral arms could create small, but important,  gravitational potential enhancements, aiding fragmentation.  Additional effects, such as the momentum exchange between 
the gas and the rocks, may have some effect as well (e.g., Johansen et al.~2007).  Moreover, a greater fraction of large solids will likely exacerbate the differences, if they can indeed be attributed to solid-gas interactions. This result is a topic by itself, and needs to be addressed in a separate study.

\subsection{SIM2: Gas and Rock Evolution}

SIM2 allows us to explore fragmentation and concentration of solids in a disk that 
is driven toward instability by mass loading.  It is not as violently unstable as SIM1, and has much more ordered spiral arms.  In addition, visible, weak spiral structure is already present at the start of the evolution presented here.  Because the particles are distributed according to gas mass, the rocks have not fully settled to the midplane by the time dense spiral arms develop, unlike SIM1.  This allows for the possibility that the spiral waves might significantly hinder midplane rock settling.

Within about 1/2 orbit at $r\sim100$ AU, 
an arm becomes very dense and fragments into a clump.  The simulation is evolved for an additional 1/2 orbit after fragmentation. At the time the simulation is stopped, a second arm is showing signs of clump formation, but we do not investigate it further here. The surface density evolution of the disk is given in Figure 13.  As done for SIM1,  the same surface density snapshots are shown in Figure 14, but with the particles superimposed.  The rocks respond quickly to the density perturbations, and the spiral arms become enhanced despite the short duration of the simulation, as highlighted in Figure 15.  Such rapid concentration indicates that 
enhancement of solids will occur as soon as spiral arms form.  Even with large-scale spiral arms, significant settling to the midplane takes place.  Within one orbit, the midplane has a very high solids concentration, with $>95\%$ of rocks mass contained  in the midplane cells.  In some spiral arms, the rock density dominates over the gas density.

In Figure 16, we show the azimuthally averaged rock surface density profile for several snapshots.  Particles move inward, but are thrown back out to larger radii as the spiral arms increase in amplitude.   As seen in SIM1, rocks are concentrated in azimuthally averaged density maxima.  In contrast to the 10 cm-size particles, the solids in SIM2km do not collect in the spiral gaseous spiral arms (Fig.~17). Spiral arms do form in the solids, but because the 1 km-size solids respond to gas drag slowly, the particles do not collect in  the high-density gas regions. This, in turn, allows for super nebular solids-to-gas ratios outside of the gaseous spiral arms (Fig. 18).  The enhancement is partly due to the much broader solid spiral arms, causing enhancement in the low-density regions immediately before and after a gaseous arm, and partly due to misalignments between the gaseous and solid arms.

\section{Clumps}

One of the principal science questions that we want to address in this study is whether clumps that form by disk instability can be enriched at birth.  As shown in the previous sections, spiral arms sweep up rocks rapidly, and whenever fragmentation does occur, clumps are born in these rock-rich environments. In this section, we address the magnitude of this enrichment and the consequences it has for gas giant composition and core formation. 

Consider first the clump that forms in SIM2 (SIM2:C1) because it has a relatively clean formation environment, i.e., it forms from one spiral arm and does not suffer mergers or collisions with other arms or clumps.  Figure 19 shows the rock-to-gas density ratio $f_{\rm RtoG}$ for the full surface density (left) and for just the midplane cells (right).  The center of the clump ($\phi\sim5.3$ rad) shows a large rock enhancement, as do the spiral arms/wakes that continue to supply solids and gas to the clump.  The dearth of rocks at larger azimuth (prograde direction;  see Fig.~14) is not completely understood, but seems to be related to formation of the nearby structure.

 In order to quantify the rock enrichment, we need an estimate for the clump mass.  Figure 20 plots cell gas volume density and the gas temperature as a function of radius from the densest cell in the clump.   The minor peaks in the profiles owe to the highly flattened clump structure, yielding poor vertical resolution after collapse.  Based on these profiles, we consider a given cell's mass to belong to the clump if its gas density $\rho\gtrsim 9\times 10^{-13}$ g cm$^{-13}$ and its temperature $T>34$ K.   
Although subjective, these cutoffs provide a reasonable definition for the clump, and
are chosen to be low, but still able to distinguish the clump from spiral arm material.  The clump masses vary by a few percent when varying the cutoff temperature by a few Kelvin or the density by $\sim10\%$. Figure 21 shows a surface density+rocks snapshot of the clump at 1300 yr (right) and just before fragmentation at 960 yr (left).  If a particle is located in a cell that belongs to a clump, as defined above, we consider the particle to also belong to that clump.  This criterion will not work in general, but for the 10 cm particles followed here, the criterion is reasonable.   Particles that belong to the clump are plotted using blue crosses, and seem to delineate the clump's extent well.  These particles are also shown on the 960 yr panel, which demonstrates that the clump forms from a 20-30 AU section of the arm, consistent with Durisen et al.~(2008) and Boley et al.~(2010), with some gas and rock mass growth after its initial formation.   
The clump at this stage of its formation is deformed, and although the solids are not concentrated near the geometric center of the clump, they are concentrated at the clump's density maximum.  

Table 1 gives the gas and rock mass of SIM2:C1 at 1300, 1450, and 1620 yr.  We also show mass estimates for gas with $\rho\gtrsim \rho_{\rm max}/2$ and for $\rho\gtrsim 0.9 \rho_{\rm max}$. These additional cutoffs allow us to explore the rock distribution in the clump.  Over 320 yr, the clump grows from 7.1 to 8.5 $M_J$.  The rock mass also increases from 35 to 42 $M_{\oplus}$, 
keeping $f_{\rm RtoG}\sim0.015$ throughout the clump's evolution.   When taking into account the entrained dust distribution, the overall enrichment of the clump $f_{\rm enr}\equiv (f_{RtoG}+0.01)/0.02 \sim1.3$.  Despite forming from the spiral arm, the enrichment of the entire 
clump is fairly modest, about 30\% over the nebula's average value.  We can understand this 
result as follows:  Although the center of the spiral arms are highly rock-enriched, the arm material  immediately surrounding the midplane of the spiral arm is rock depleted.  Collapse mixes this material back together, resulting in a smaller enrichment than what one might expect from the densest regions of the spiral arm alone.  Mass growth does not appear to change this ratio significantly.  This is likely due to growth being mediated by filaments (see Fig.~19), which are rock-enriched. 

Even though the clump as a whole has modest enrichment, the rocks are highly concentrated near the center.  At half the peak density, the clump is enriched by a factor of a few in the rocks and about two in total solids.  Near the center of the clump, the rock-to-gas ratio is ten times the average nebula value, with 30 to 38 $M_{\oplus}$ of rock concentrated at the center.  Even if the entire clump is eventually disrupted, extreme rock concentration could lead to the formation of early cores and planetoids in the outer disk.

In contrast to SIM2:C1, SIM2km:C1 is depleted in rocks, as km-size solids are not as easily captured as the 10 cm-size solids.  Table 1 shows that there is no additional concentration of solids toward the center, so unlike the 10 cm case, no differentiation has taken place.  Overall, the clump is subnebular in solid composition, is less massive than its 10 cm counterpart, and has a lower central density.  This provides additional evidence that the evolution of the solids affect the evolution of the gaseous disk, even for features as large as fragments.  The change is not as drastic as that seen between SIM1mu and SIM1muNoRocks, but SIM2 is not as flocculent as SIM1, either.  

The low level of enrichment cannot be attributed to lack of interactions between solids and the gaseous fragment. Figure 22 traces the solids associated with SIM2km:C1 at about 1720 yr, determined the same way as for SIM2:C1, and traces the particles backwards and forwards in the clump's orbit.  In the earliest snapshot shown (1290 yr), roughly 1 $M_{\oplus}$ of solids have yet to enter the planet's Hill sphere, while for the oldest snapshot (2190 yr), about 0.7 $M_{\oplus}$ of the solids escape have passed out of the new, larger Hill sphere.  This does show that some planetesimals are being captured into the orbit of the protoplanet, but it does not lead to enrichment. In the latter snapshot, the clump has grown to roughly 12 $M_J$, but only has 11 $M_{\oplus}$ of large solids, giving a rock-to-gas ratio of 0.0029, which makes the protoplanet more anemic than seen for the snapshot in Table 1.  The capturing of solids for this clump appears to be due to the capture of only the lowest relative velocity solids, aided by a gradual deepening of the protoplanet's potential well due to gas accretion.

Unlike the SIM2 simulations, SIM1mu shows prodigious clump formation with very flocculent spiral structure.  Clumps merge, scatter, and interact with material arms.  These interactions sometimes lead to mass growth, while at other times, the clumps become sheared away.  For this discussion, we 
consider three clumps that have different but illustrative histories.   We denote them as SIM1mu:C1, which becomes tidally disrupted, SIM1mu:C2, which grows to a brown dwarf mass scale, and SIM1mu:C3, which stays a high-mass gas giant (Tables 2 and 3).  
The same density cutoff as used for SIM2:C1 provides a reasonable differentiation between clump and arm material, but a slightly higher temperature threshold ($T>40$ K) is chosen.   

SIM1mu:C1 forms between 940 and 1100 y, but only survives about 1/2 of an orbit before it migrates inward and becomes tidally disrupted.  Its density and temperature profiles are not shown, but its mass is given in Table 2. 
Even though the clump becomes disrupted, it is worth noting that the central concentration in SIM1mu:C1 is 27 times the nebula's average rock-to-gas ratio, placing 38 $M_{\oplus}$ of rock at the fragment's center.  Unfortunately, we do not have the resolution to follow the formation of these rocks into a self-bound core, but the results of  Helled \& Schubert (2008) do suggest 
that formation of such a core is quite possible (see also Boss 1998).  In particular, our clumps start differentiated, 
whereas Helled \& Schubert assume homogenous composition, so the actual timescale for core formation may be 
shorter than found in the Helled \& Schubert simulations.

The second clump to form in SIM1mu, SIM1mu:C2, survives until the end of the simulation.  Its temperature and density profiles for several snapshots are given in Figure 23. The clump shows prodigious mass growth from 7 to 32 $M_J$, mainly in large bursts due to mergers with other clumps and from collisions with material arms.  Even with its violent history and rapid mass growth, its total enrichment factor remains $f_{\rm enr}\sim1.4$ times the nebula's average value.  This is likely due to its amalgamation with other rock-enriched structures, keeping $f_{\rm RtoG}$ roughly constant.  As with SIM1mu:C1, the rocks are highly concentrated and could form a large core.   How such initial differentiation would affect early brown dwarf evolution observationally is unclear. 

Finally, SIM1mu:C3 also survives until the end of the simulation, with a mass that hovers around 10 $M_J$.  Unlike SIM1mu:C3, it does not merge with another clump or pass through a material arm.  Its $f_{\rm enr}\sim1.4$, 
and it has a highly concentrated rock core.    

\section{Discussion}

Fragmentation at large disk radii should occur for some systems during the mass loading phase of disk evolution, i.e., while there is still significant infall onto the disk.  Formation by disk instability need not exclude formation by core accretion, and both formation channels could operate in the same system.  Although the two main formation mechanisms may be discernible through statistical trends, as discussed in the introduction, orbital migration and/or scattering will make it difficult to attribute a formation mechanism to an individual planet.  Additional observational signatures are therefore desirable, and chemical and structural (cores) variations may allow for such differentiation.  Building on the work of, e.g., Haghighipour \& Boss (2003), Rice et al.~(2004, 2006), Johansen et al.~(2007), and Paardekooper (2007), we have included solids in our three-dimensional hydrodynamics code to investigate how the distribution of solids, particularly within clumps, is affected by gravitational instabilities. 

\subsection{Spiral Arm-Solids Coupling}

GIs form dense spiral structures that are very efficient at concentrating solids with sizes $\sim 10$ cm, with surface density enhancements of solids 30 times larger than the average nebula value, in agreement with the Rice et al.~(2004) study.   Near the midplane, the mass density can become dominated by 10 cm rocks.  Super rock-to-gas ratios even occur in SIM2km due to the spiral arms of solids being offset and broader than the gaseous arms.  We also find that solids do not simply drift inward; instead, they follow the evolution of the gas spiral structure, including outward migration.  Simulations should be extended over much longer periods than studied here so that the asymptotic behavior of GIs and solids can be addressed, but these results do show that solids need not be lost by drag and that large particles could be kept at large radii while GIs operate.  In addition, the results of SIM1muNoRocks  suggest that solids can affect the gas evolution.  As discussed earlier, this result requires further study, as numerical chaos could be contributing to some of the differences that we see between SIM1muNoRocks and SIM1mu.  However, SIM2km also showed differences in the evolutions of clumps SIM2:C1 and  SIM2km:C1, indicating that more than just numerical chaos is at work.

\subsection{Clump Enrichment}

Clump formation takes place exactly where rock-size solids are concentrated, i.e., the spiral arms.  Although the midplane regions of these arms can be highly enriched, disk instability only leads to modest enrichment within the fragments, with a total enrichment of tens of percent over the nebula's average value.  This value may change for different rock distributions, but we argue that these results provide an upper limit on the amount of  enrichment in fragments, as spiral arms are very efficient at sweeping up 10 cm rocks for the conditions explored here. This is emphasized by SIM2km, which actually shows a depletion in solids relative to the gas mass.  However, Figure 18 also shows that there are supernebular solids near the clump for the 1 km conditions.  If solid evolution is considered, then some of the depletion seen in SIM2km may be marginalized.  Using the limits that the 10 cm and 1 km simulations provide, we can provide an upper limit to the amount of enrichment at birth, assuming rock concentration does not become even more efficient for larger metallicities or for different opacity/large solids fractions. If we assume that most of the solids grow to 10 cm sizes and that the rock enrichment seen in these simulations is doubled as a consequence, we would expect to have a typical $f_{\rm RtoG}\sim0.03$, giving a total enrichment $f_{\rm enr}\sim 2$.  This suggests that a disk instability planet, although enriched, will have at birth no more than twice the amount of metals compared with its host star, taking our results at face value. If a substantial amount of the protoplanet's envelope is tidally stripped, the final planet could become more enriched than estimated from the initial clump enrichment (see section 6.4). 

Enrichment at birth provides a base level for other enrichment mechanisms, such as planetesimal capture.  However, Helled \& Bodenheimer (2010) found that enrichment by planetesimals with sizes $\ge1$ km is negligible at radii of $r\sim70$ AU for the conditions they studied.  Because fragmentation is expected to be much more likely at large radii (e.g., Stamatellos et al.~2007; Stamatellos \& Whitworth 2008; Boley 2009; Rafikov 2009; Clarke 2009), we expect enrichment at birth to be the dominant enrichment mechanism.  Nevertheless, if rapid migration occurs before dissociative collapse or if fragmentation occurs, however rarely, at 10s of AU, then planetesimal capture could also enhance disk instability objects.  

\subsection{Opacity and Grain Growth}

Grain growth alters the opacity of the disk (D'Alessio et al.~2001) and changes the cooling rates, which in turn affects the behavior of GIs  (Cai et al.~2006, 2008).   Our results are consistent with 
this picture, where fragmentation is suppressed in SIM1mm due to the higher opacities.  The problem is, unfortunately, complicated by the competing effects of grain growth, which suppresses fragmentation at large radii, and settling, which would reduce the opacity for a large volume of gas.  This emphasizes that, to address this problem properly, 
grain growth models need to be included with 
the dynamical evolution of solids in order to 
model fragmentation of gaseous disks correctly.   Concerning grain growth, we have assumed for this study that 10 cm or 1 km particles are present by the time the disk fragments.  Whether this can be achieved is unclear, but the observations of Greaves et al.~(2008) do suggest that 10 cm particles are present in massive, extended disks.

\subsection{Clump Cores and Fates}

We have assumed that the clump masses calculated here are comparable to the final gas giant/brown dwarf masses.  Based on the arguments presented in Boley et al.~(2010), we do not expect rapid mass growth after a clump undergoes its second collapse due to dissociation of H$_2$ (the fragmentation process being the first collapse), which should occur after a few thousand years for these clump masses (see also Helled \& Schubert 2008).  Because the centrifugal radii of gas fragments are expected to be a few tenths to an AU at disk radii $\sim 100$ AU, a substantial disk will likely form as the fragment collapses.  Subsequent mass growth will be mediated by subdisk evolution, which has recently been demonstrated in simulations by Vorobyov \& Basu (2010).   Moreover, the clump masses calculated here represent objects that will become 
either gas giants  or brown dwarfs with surrounding disks, where the disks themselves may be massive.
Global instabilities in the clumps, e.g., the bar instability (Durisen et al.~1986), could also redistribute angular momentum before dissociative collapse, ejecting tens of percent of the clump's mass.  
If convection is unable to redistribute the centrally concentrated rocks before collapse, then such instabilities would preferentially eject the  solid-poor material.  However, even this scenario would only increase the enrichment of the clump by tens of percent.  Only if most of the clump's initial mass could be stripped, e.g., due to tides, would the enrichment of the clump match that of Jupiter's ($ f_{\rm enr}$ up to 6; see, e.g., Baraffe et al.~2010).  

Tidal disruption of clumps with dense rock concentrations is seen in these simulations. For example, SIM1mu:C1 is an 8 $M_J$ clump with 38 $M_{\oplus}$ of solids at its center, and becomes sheared away within half an orbit.  Unfortunately, we do not have the resolution to follow core formation, so investigating 
whether these solid concentrations can separate from the gas (see, e.g., Rice et al.~2006) and form large cores will need to be addressed in a future study.  However, it is tempting to speculate whether a third formation channel for gas giants, which we call
``core assist plus gas capture,'' becomes available if cores do form in clumps that eventually become disrupted.  
In this mechanism, gas fragments behave as factories for the formation of rocky/icy cores.  After fragments become tidally disrupted, their cores could be liberated into the disk and begin to capture gas slowly, possibly even in the later and less active phase of disk evolution. 
This is similar to other hybrid gas giant formation mechanisms where hydrodynamical instabilities accelerate core formation by concentrating solids (e.g., Durisen et al. 2005, Klahr and Bodenheimer 2006).    We also acknowledge that core formation through differentiation at birth may be subject to complicated equation of state effects as the pressure and density increase as the clump contracts. 

As we have shown here with the 1 mm opacity simulation, fragmentation is very sensitive to opacity, and dense knots or short-lived clumps could be more common than long-lived fragments (see, e.g., Vorobyov \& Basu 2010). In this case, core assist plus gas capture may very well represent another channel for {\it in situ} formation of gas giants on wide orbits, if the gravitational torques and gas interactions keep the cores at large radii.  If, instead, the cores migrate into the inner nebula as they capture gas, they will produce inner gas giants.  Interestingly, we note that HR149026b has presented a puzzle for planet formation due to its extremely large core estimate ($\sim70 M_{\oplus}$; Sato et al.~2005).  If SIM1mu:C3 were to be sheared away, its concentration of $70 M_{\oplus}$ of solids in its central regions could give rise to the formation of such a planet by core assist plus gas capture. 

\subsection{Formation Channels For Substellar Companions}


Our results suggest that there may be three formation channels for gas giants planets (core accretion plus gas capture, direct formation by disk instability, and core assist plus gas capture), which could give rise to three populations of gas giants.  If this is indeed the case,  it is crucial to understand how the formation channels could be distinguished observationally.   
For this discussion, let us refer to a population I gas giant (Pop I GG) as one that forms via traditional core accretion plus gas capture, population II GGs as disk instability gas giants, and pop III GGs as core assist plus gas capture.

If pop I GGs are typically as enriched as Jupiter and Saturn, then a metallicity difference between pop II and pop I GGs may indeed be observable. Unfortunately, the typical outcome of core accretion has not yet been established, as core accretion simulations follow planet formation up until the runaway gas accretion phase is established, but not the end of accretion and subsequent evolution (e.g., see recent work by Movshovitz et al.~2010).  As a result, the final metallicity of pop I GGs  is unclear.  To illustrate this point, take the enrichment of heavy elements in Jupiter to be about five times the solar value.  If Jupiter had continued to accrete gas with nearly solar composition, it would need to grow to about 4 $M_J$ before its enrichment would drop to twice solar and becoming indistinguishable, using metallicity, from pop II GGs.   Looking for trends in the frequency of GGs as functions of semi-major axis, host star metallicity, and time, as discussed in the introduction, may still be the best way to distinguish pop I and II GGs.  

Pop III GGs suffer the same uncertainty in final metallicity as do pop I GGs.    In the simulations 
presented here, we unfortunately cannot 
compute the formation of a solid core,  let alone the subsequent runaway gas capture onto that core, and  cannot state what the final enrichment of such an object will be.  We do note however that, with the large concentration of solids seen in gas clumps, Pop III GGs may be very similar to Pop I GGs, but have very large cores for their mass (e.g., HR149026b).  Whether Pop II or III objects can form 
in a given disk will likely depend on the protostellar core's angular momentum.  The fraction of  protostars
for which Pop II and III GG formation will be important is uncertain, but 
we remind the reader, as also noted by Stamatellos \& Whitworth (2007), that the distribution of angular momentum within dark cores 
probably does permit the formation of extended disks (Goodman et al.~1993).  


\section{Conclusions}

We have investigated the coupling between rocks and gas in fragmenting, gravitationally 
unstable disks.  We have explored the behavior of 10 cm-size and 1 km-size particles in this study.  The former corresponds to sizes that are expected to be swept up efficiently by spiral arms, while the latter is largely decoupled from gas drag.  Because the small-grain limit is simply entrained with the gas, these simulations explore the two nontrivial regimes for the behavior of solids in gravitationally unstable disks.  Our simulations have many implications for disk fragmentation and planet formation, which we summarize in bullet form.

\begin{itemize}

\item Spiral arms concentrate rock-size material into the regions that are most likely to fragment.  However, this aerodynamic capturing of solids leads to, at best, a total enrichment of a clump by a factor of two.  A distribution of solids that does not contain a large amount of rock-size particles will reduce this effect, without considering solid evolution, as demonstrated by the 1 km-size run.
\item Subsequent mass growth does not dilute the birth enrichment in these simulations for the 10 cm-size solids because clumps accrete gas from filaments, which are rock enriched, or grow by collisions with spiral arms and/or other clumps, which are also rock enriched.
\item If most of the solid mass is in 1 km-size solids or in sizes that are equally decoupled from the gas, clumps can become anemic in heavy elements.
\item Rock densities can become comparable to or exceed local gas densities in some structures. Even for the 1 km-size case, regions of supernebular solids-to-gas occur due to differences between the spiral structures in the solids and the gas. 
\item The rock distribution appears to affect the behavior of gravitational instabilities and fragmentation.  The influence of the solids on the gaseous disk, particularly fragmentation, may mainly be due to gravitational potential changes.  Additional simulations with distributions of solids need to be explored, but must be coupled with a solid size evolution model. 
\item Gravitational instabilities interfere with the simple inward drift of solids.  Instead, rocks that would normally have a very high inward velocity are entrained with the evolution of the spiral arms, and are even transported out to large radii.
\item The fragmentation process is affected by the grain size distribution, i.e., the opacity. For the temperatures in the outer disk, larger grain sizes can stop dense condensations from forming long-lived clumps.
\item Although the total enrichment is modest, solids are highly concentrated near the center of the clump as part of the fragmentation process.  The solids-to-gas ratio can be as high as a few tenths in these regions, with tens of Earth masses of rock available for immediate core formation, i.e., the material does not need to settle from the entire clump.  This could significantly decrease the core formation timescales calculated by Helled \& Schubert.~(2008).  Overall, the results suggest that disk instability planets can have massive cores.
\item Fragments can become sheared away by tides, but because rocks are concentrated near the center of the clump, these fragments may still form rocky cores before the gas is completely disrupted. If such solid cores survive tidal disruption of the extended gas envelope, they may very well represent the first rocky/icy planets  or planetary embryos
in a planetary system.

\end{itemize}

Understanding the behavior of solids in gaseous disks is required for building a basic picture of planet formation, even in the context of a gravitationally unstable disk.  As demonstrated here, fragments, although only modestly enriched, 
can be differentiated at birth. This leaves open the possibility of a hybrid scenario that we call core assist plus gas capture.  In addition, it also suggests that clumps as massive as brown dwarfs could also start as differentiated objects, depending on the degree of the enhancement of solids within the spiral arms at the time of gas disk fragmentation.   These formation channels should have observational consequences, which we only begin to touch upon here, but include metallicity and core size, as well as orbital locations if done 
in a statistical sense. 
Because the fragmentation process is strongly affected by dust characteristics, future studies should include  grain growth models that self-consistently change dust opacities and capture, even if heuristically, the formation of solids with a distribution of sizes.   Although challenging, such a model will greatly aid in determining the likelihood of fragmentation and what type of objects disk instability will typically form. 

ACB and RHD would like to acknowledge the following individuals who have influenced our thinking and have provided us with ideas regarding the topics discussed in this manuscript: 
F.~Adams, P.~Bodenheimer, A.~Boss, C.~Dullemond,    E.~Ford,  L.~Fouchet, N.~Haghighipour,  L.~Hartmann,   T.~Hayfield,  R.~Helled, Th.~Henning, O.~Hubickyj,  A.~Johansen, H.~Klahr, J.~Lissauer, G.~Lodato,   L.~Mayer, S.~Michael,  F.~Miniati,  G.~Morfill, M.~Payne, K.~Rice,   and D.~Stamatellos.  We would particularly like to acknowledge the hospitality of the Max Planck Institute for Astronomy and 
the Max Planck Institute for Extraterrestrial Physics over the years.  ACB would like to thank the Kavli Institute for its hospitality during the ``Theory and Observations of Exoplanets''  workshop.   ACB's support was provided in part by a Theory Fellowship through the University of Florida and in part under contract with the California Institute of Technology (Caltech) funded by NASA through the Sagan Fellowship Program. RHD was supported by NASA Origins of Solar Systems grant NNX08AK36G. This research
was supported in part by the National Science Foundation under Grant No.~PHY05-51164. Resources supporting this work were provided by the NASA High-End Computing (HEC) Program through the NASA Advanced Supercomputing (NAS) Division at Ames Research Center.

\appendix
\section{Drift Tests}

A difficulty with our drag algorithm  is that it only includes the drag force when calculating the new $\delta v$.  If the stopping time is much shorter than the hydrodynamic step, then the asymptotic behavior, e.g., radial drift velocities, could be poorly calculated.  In this section, we present a simple test that addresses this issue.  We find that the asymptotic solution is affected by our approximation, but for the regime studied in the simulations presented above, our approach is valid.

Consider equation (1), but in the limit that the particles are subsonic, i.e., $f_e\approx 1$.  Take the gas velocity to be zero, so the velocity difference represents just the velocity of the particle.  Assume that gravity, $g$, is constant and that the particle starts at rest  to find 
$v_d = g t'_s(1 -\exp(-t/t'_s))$,
where $t'_s=(\pi/8)^{1/2} t_s$. The asymptotic drift velocity is simply $g t'_s$.  In the context of a disk, we can compute the asymptotic settling velocity of a particle by $v_z=-\Omega^2 z t'_s$, assuming the stopping length is small. For the radial velocity, we follow Weidenschilling  (1977) and consider the radial drift regime, where $t'_s$ is small compared with the orbital time, as well as the perturbed orbit regime, where $t'_s$ is long compared with the orbital time.  For the radial drift regime, $v_r= (\Omega_{\rm gas}^2- \Omega_K^2)r t'_s$, and for the perturbed orbit regime, $v_r= (\Omega_{\rm gas}^2- \Omega_K^2) r /(t'_s\Omega_K^2)$.  

For this test, we envisage a disk that has $T=30$ K and $\rho=10^{-13}$ g cm$^{-3}$, everywhere.  The central star is assumed to have a mass of 1.5 $M_{\odot}$, but the gas is fixed to orbit as if the star's mass were 1.35 $M_{\odot}$.  The radial and vertical motions of the gas are assumed to be zero.  For particles 10 cm in size or smaller, the particle is placed in the disk at an initial $r=100.2$ AU and $z=4.2$ AU with an $\Omega_{\rm initial}=\Omega_{1.5 M{\odot}}$.  At sizes of a meter or larger, the particles need more time to adjust to the asymptotic solution, so they are placed at $r=200.2$ AU and $z=4.2$ AU.  
For the drift speeds, we measure $v_r$ and $v_z$ as the particles cross $r=95$ AU.  
We use a time step $\sim 0.03$ yr, which is similar to the hydrodynamic time step in SIM1 and SIM2.  

We test 
particles that are 100 m, 10 m, 1 m, 10 cm, 1 cm, 1 mm, and 100 $\micron$ in radius, each with an internal particle density $\rho_a=3$ g cm$^{-3}$.    The results are shown in Figure 24. The general behavior is quite good through several orders of magnitude.  The 100 $\mu$m case does show stronger coupling than expected, and we attribute this to our assumption that only the drag force matters during the drag update.  For the vertical settling speeds, we test particles sizes of $100\micron$, 1 mm, and 1 cm, for which we find the ratio between the actual and expected settling to be 0.94, 0.99, and 1.01, repsectively.  Larger particles are not included in the settling test because they are not expected to reach the asymptotic solution.  For example, the 10 cm particle has a $t'_s\sim 200$ yr, which is about 1/4 of the orbital period at $r\sim100$ AU.  For the disk parameters in these simulations, our drag algorithm 
seems to be sufficiently accurate.  The reader should also note that, in these regions of disks, 10 cm particles have the largest drift velocities, which is why we chose to explore this size scale.


\begin{figure}
\includegraphics[width=8cm]{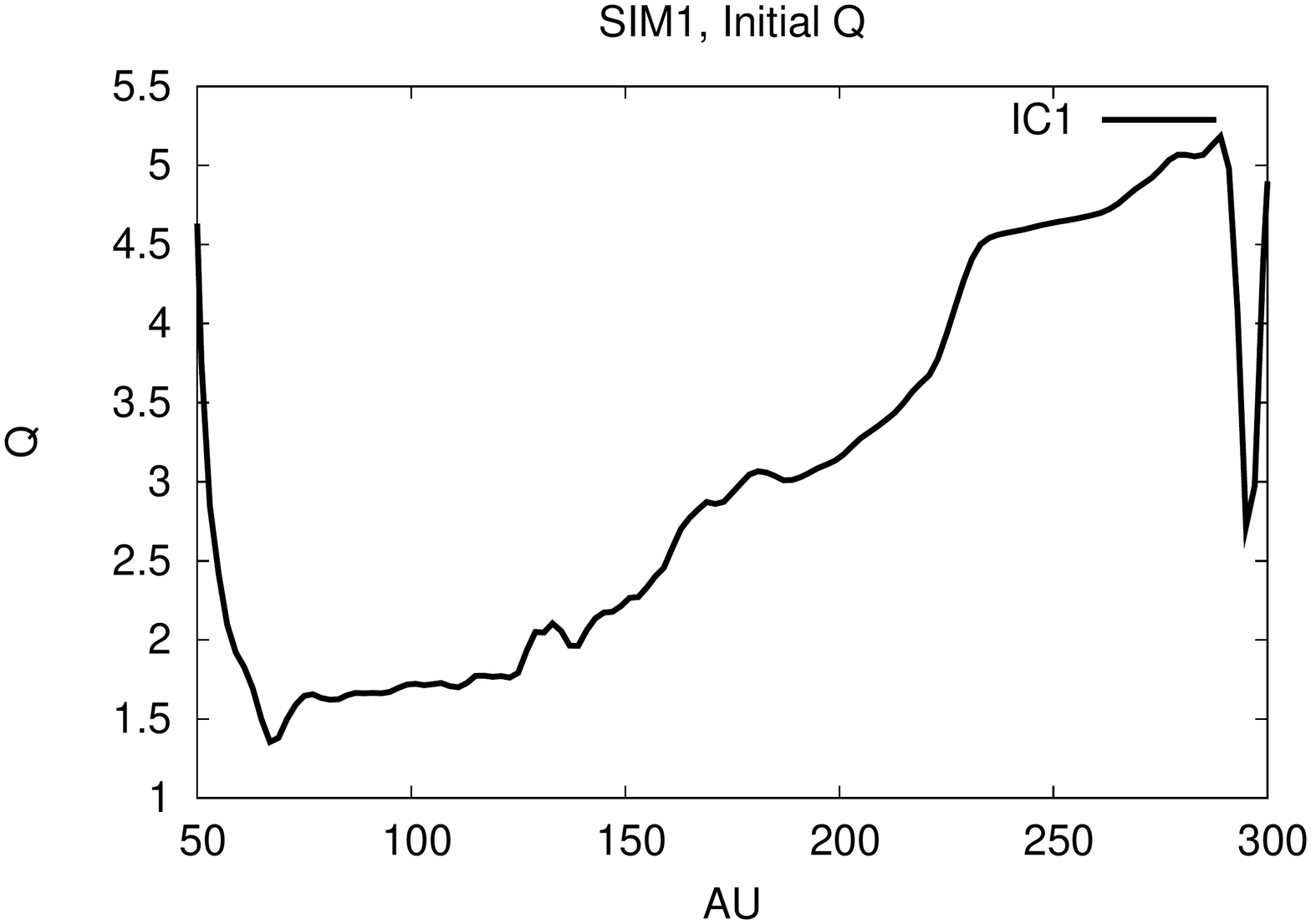}\includegraphics[width=8cm]{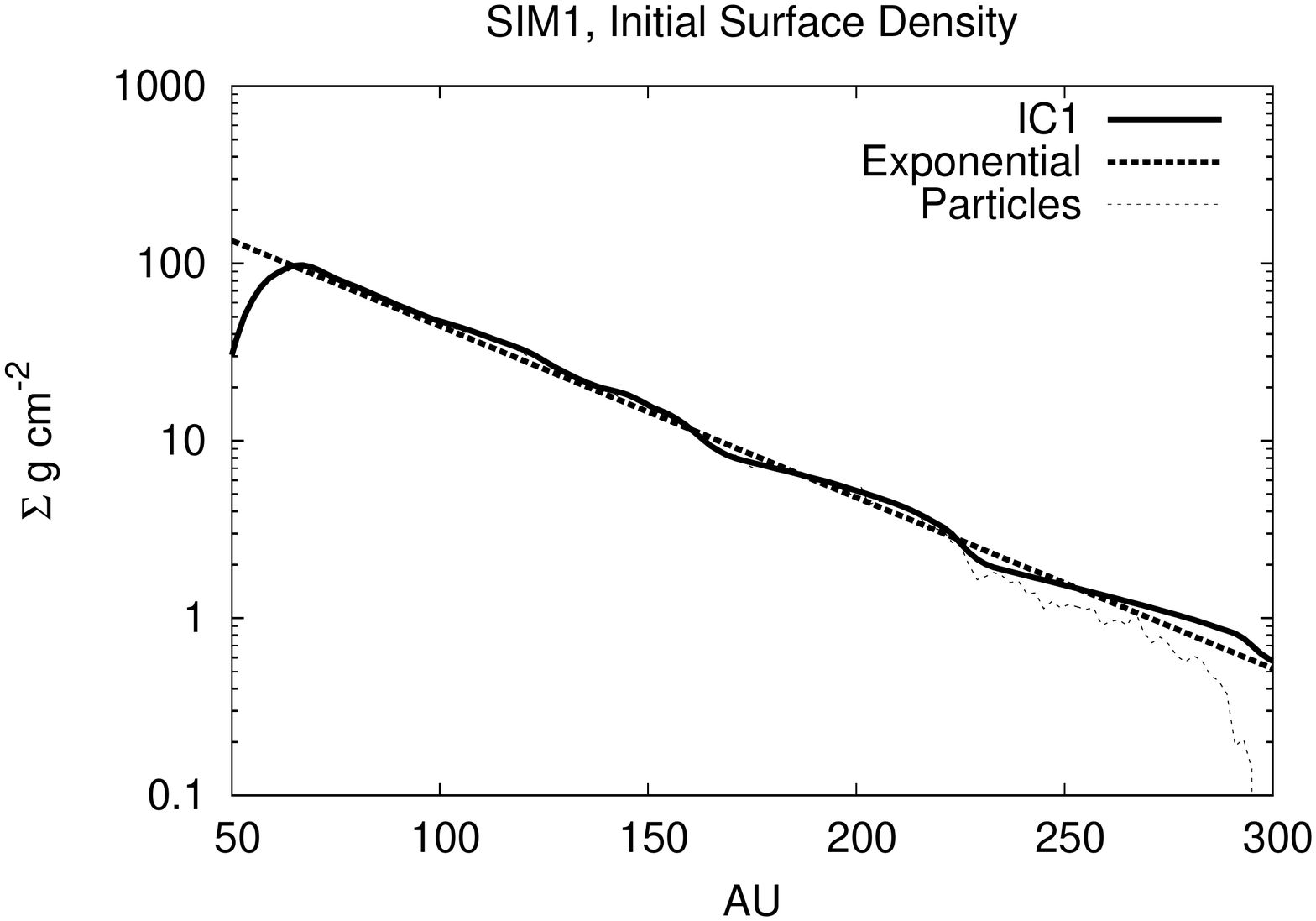}
\caption{Initial 
Toomre $Q$ and 
surface density $\Sigma$ profiles for the ICs for SIM1. The particle surface density is multiplied by a factor of 100 in this plot. }
\end{figure}

\begin{figure}
\includegraphics[width=15cm]{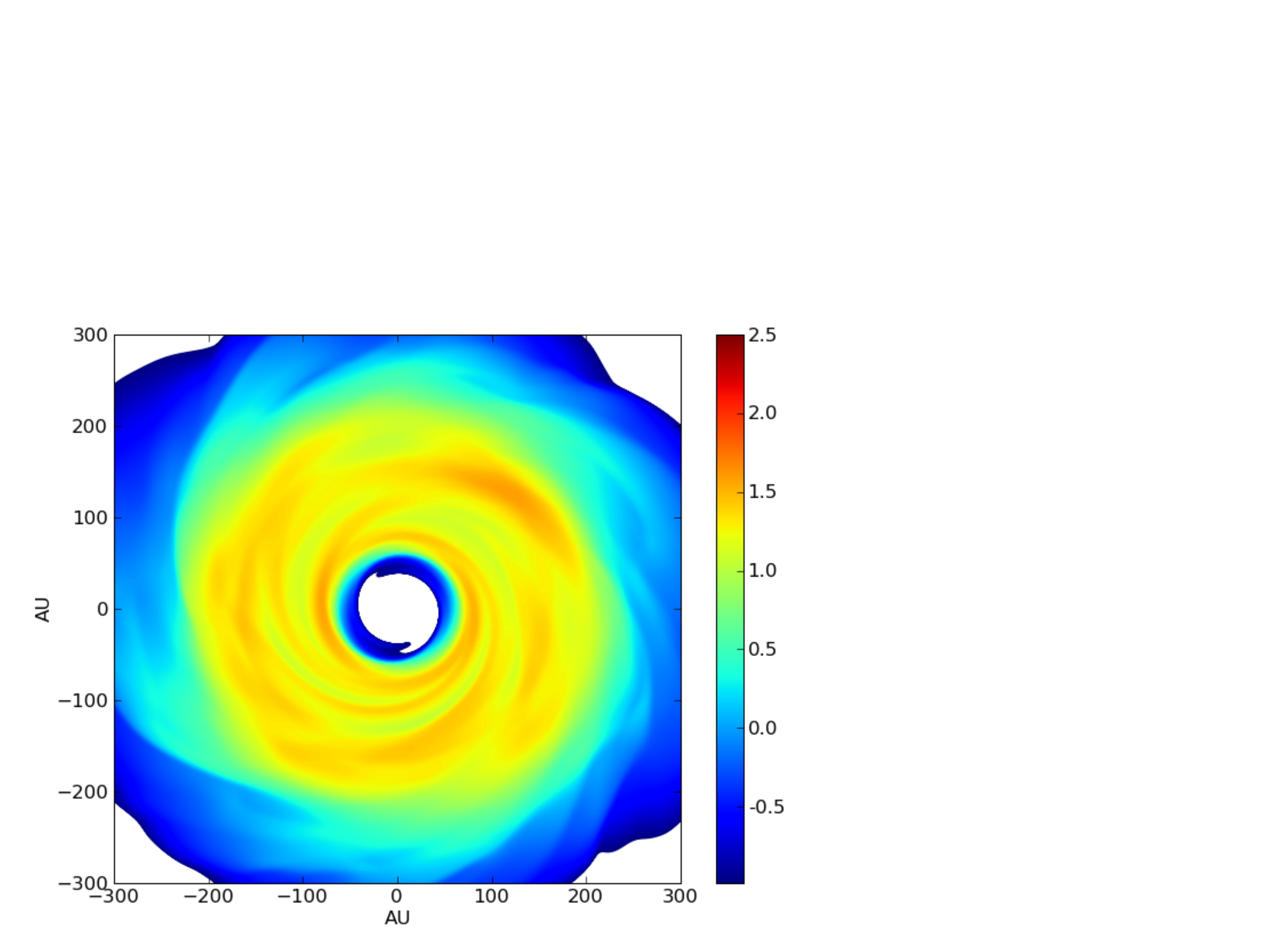}
\caption{Face-on surface density of the ICs for SIM2 .  The colorbar shows $\log \Sigma {\rm (cgs)}$. Spiral waves are clearly forming, and in about 1/2 of an orbit, the over density near $x,y=150,150$ AU will form a dense spiral arm and fragment.}
\end{figure}

\begin{figure}
\includegraphics[width=8cm]{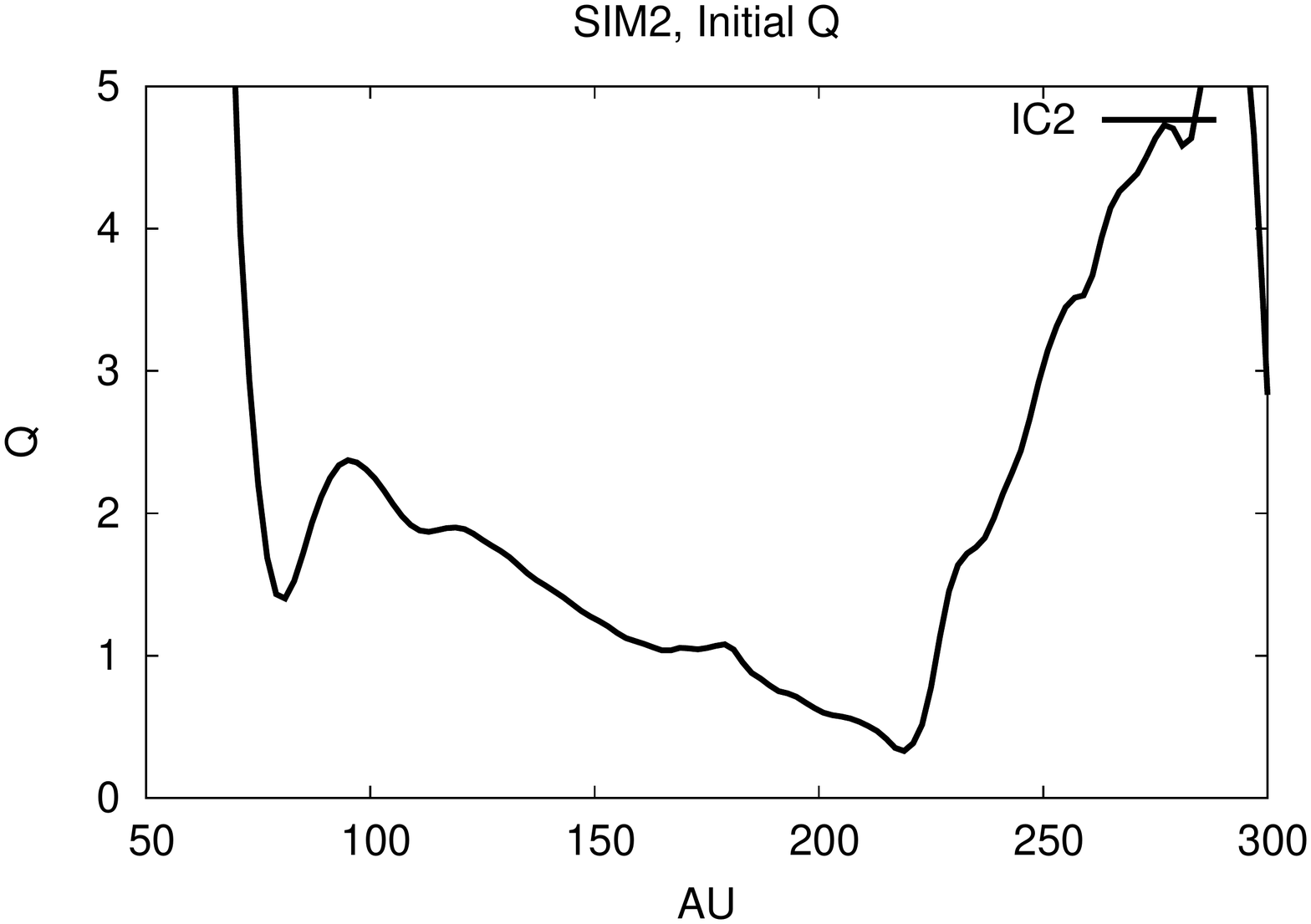}\includegraphics[width=8cm]{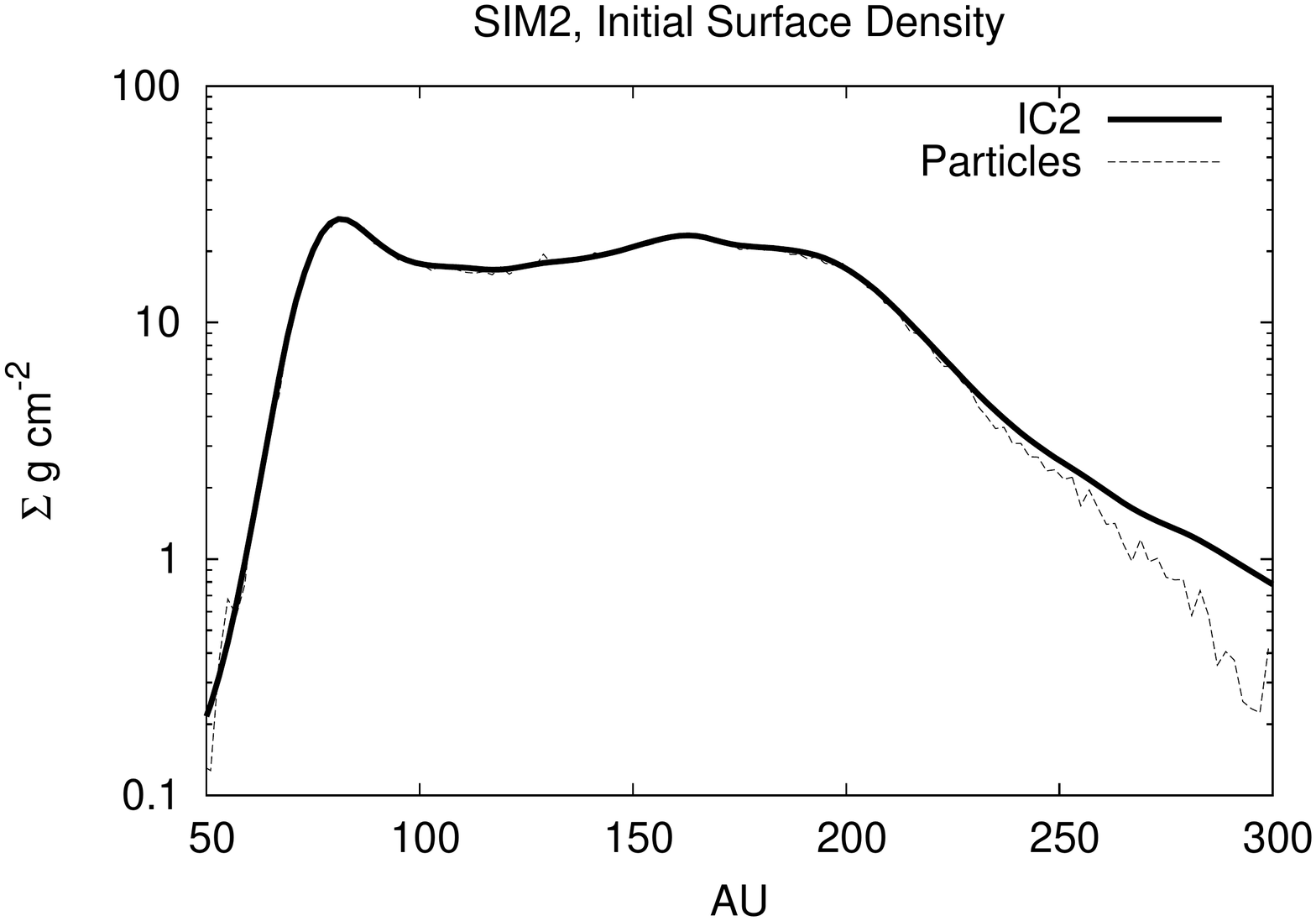}
\caption{Initial $Q$ and $\Sigma$ profiles for the SIM2 ICs.  The shallow infall profile causes mass to build up at large radii during disk formation, driving the outer disk toward extreme instability. The particle surface density is multiplied by a factor of 100 in this plot.}
\end{figure}


\clearpage

\begin{figure}
\includegraphics[width=6cm,angle=-90]{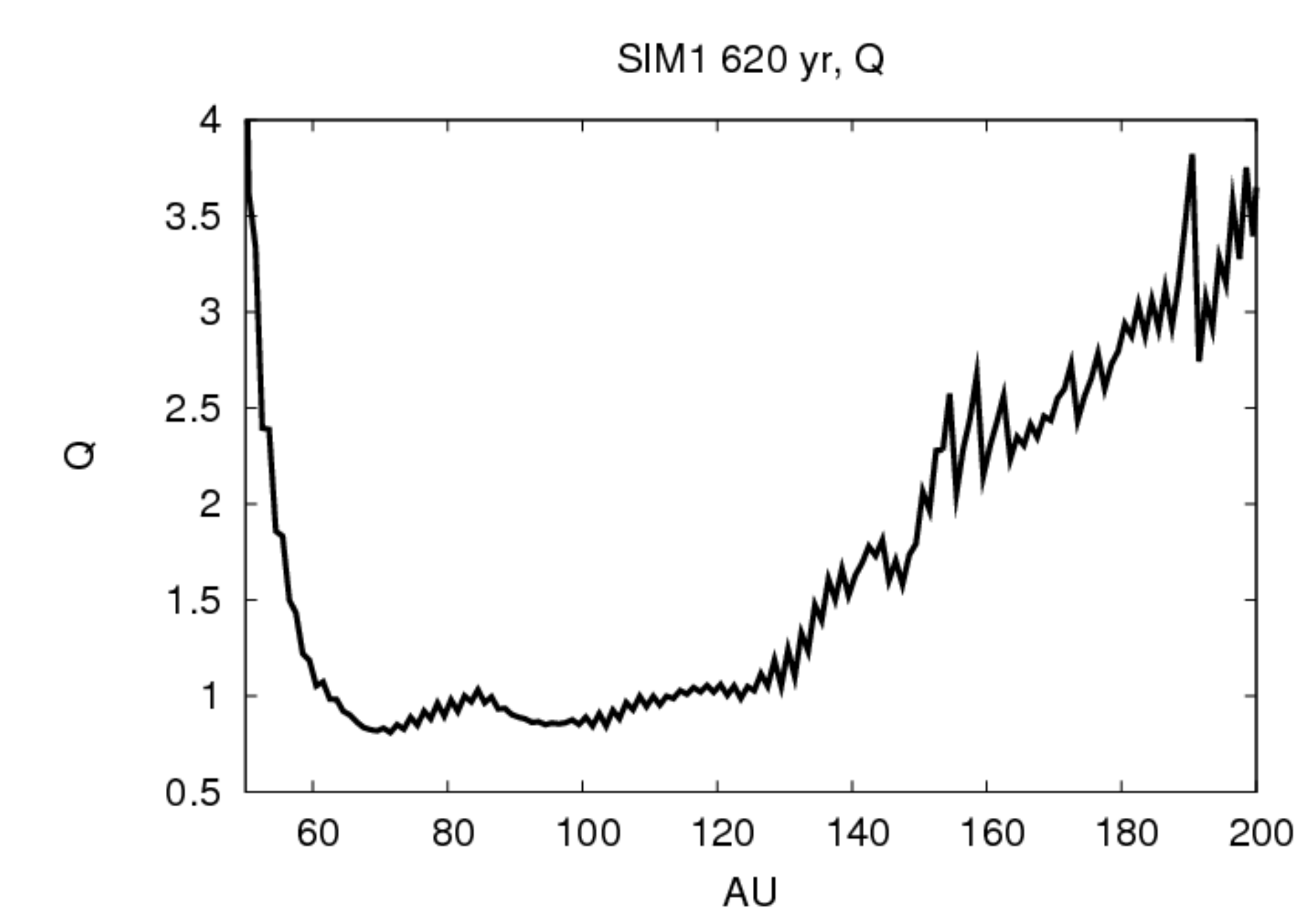}\includegraphics[width=6cm,angle=-90]{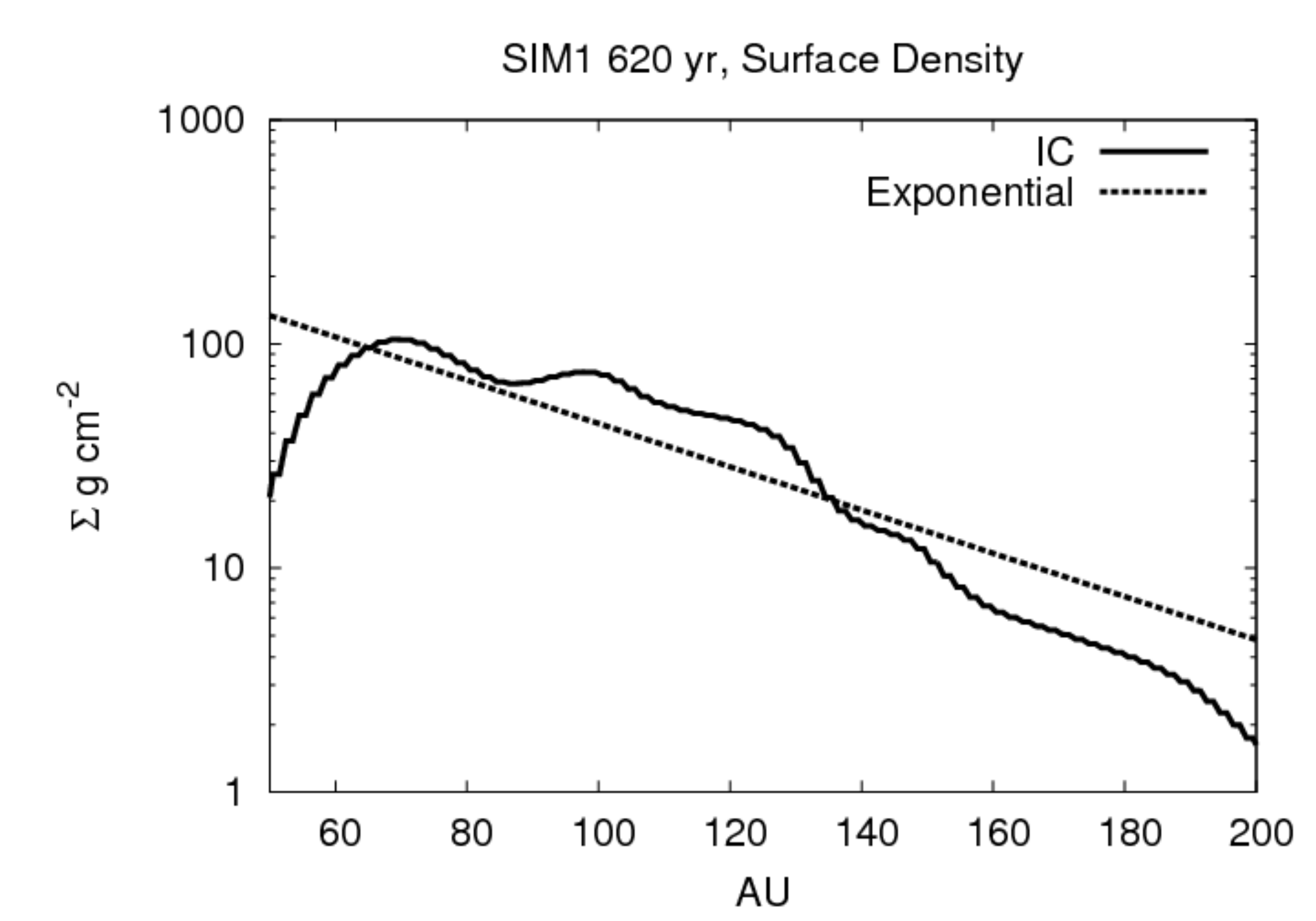}
\caption{Toomre $Q$ and $\Sigma$ profiles just after injecting the disk onto the high physical resolution grid.  The choppy $Q$ profile is a result of the direct injection method, which creates a rough epicyclic frequency profile.  Because this is only done once during the disk simulation, its effect is
just to add extra noise at this particular snapshot. }
\end{figure}

\begin{figure}
\includegraphics[width=15cm]{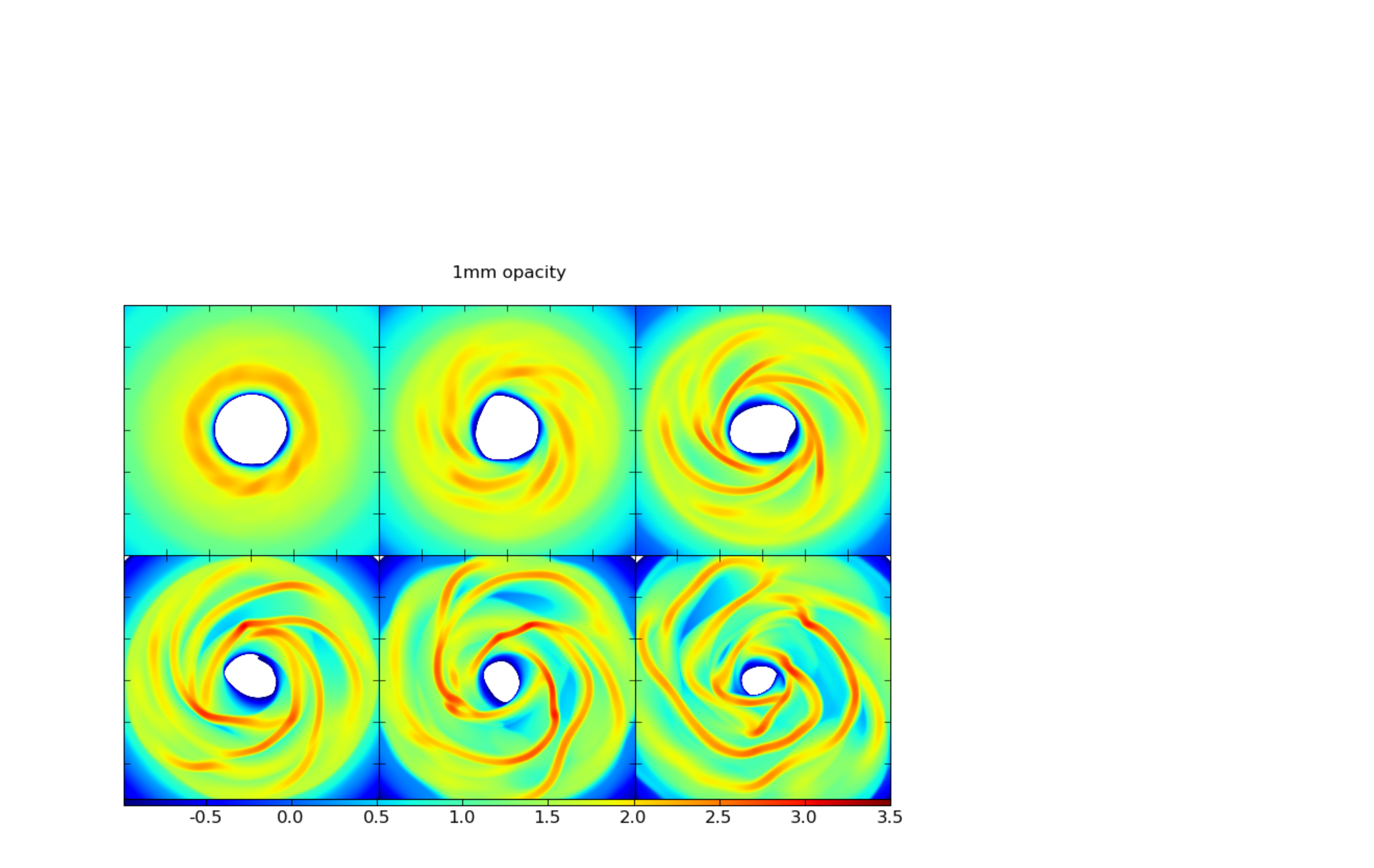}
\caption{
Gas surface density snapshots for SIM1mm. Each square is 300 AU on a side, and the snapshots correspond to  340, 620, 780, 940, 1100, and 1260 yr going from left to right and top to bottom. The colorbar shows the logarithmic surface density in  g cm$^{-2}$. }
\end{figure}

\begin{figure}
\includegraphics[width=15cm]{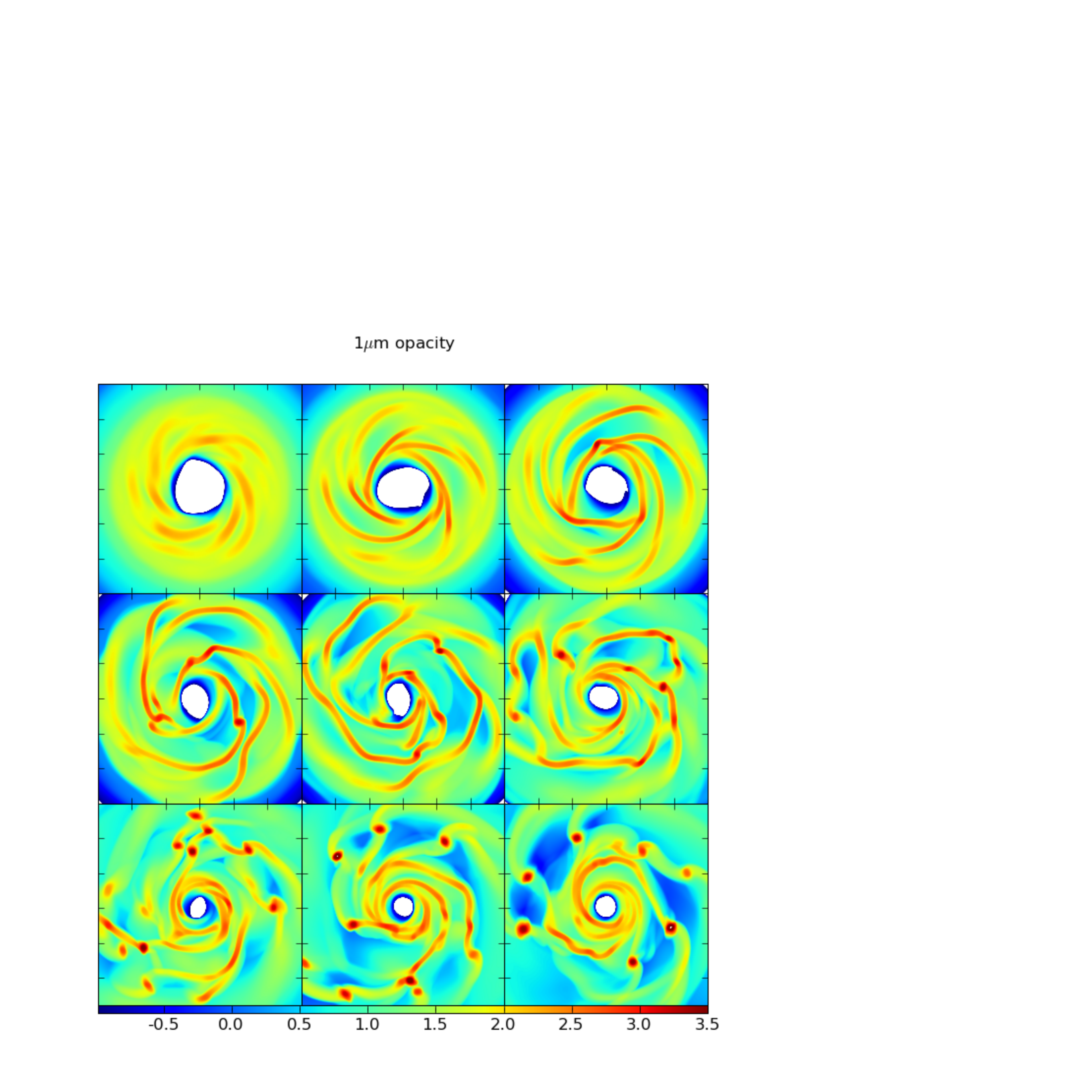}
\caption{
Gas surface density snapshots for SIM1mu. Each square is 300 AU on a side, and the snapshots correspond to 620, 780, 940, 1100, 1260, 1410, 1570, 1720, and 1870 yr going from left to right and top to bottom. The colorbar shows the logarithmic surface density in  g cm$^{-2}$.}
\end{figure}

\begin{figure}
\includegraphics[width=8cm,angle=-90]{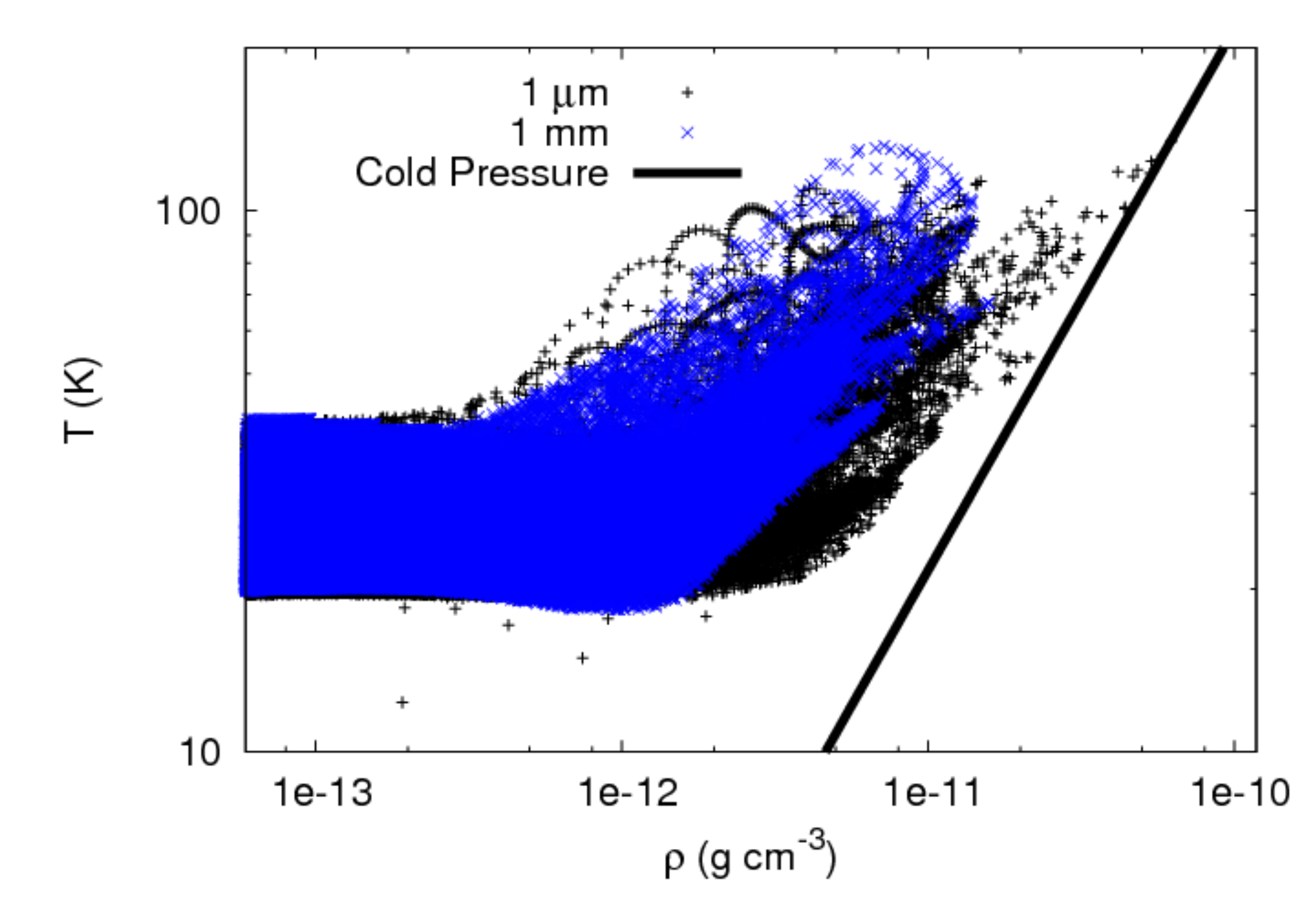}
\caption{A plot of the gas temperature versus density values at all computational grid points for the 1 mm and the 1 $\mu$m opacity simulations corresponding to the 1260 yr snapshot. For the same temperature, the 1 $\mu$m case can reach higher densities, which eventually leads to fragmentation.  The line called Cold Pressure, shows the thermodynamic temperature that would correspond to a given density for the given cold pressure.  Clump formation occurs before this threshold is met, and cannot be responsible for altering fragmentation in the 1 mm case. }
\end{figure}

\clearpage

\begin{figure}
\includegraphics[width=8cm,angle=-90]{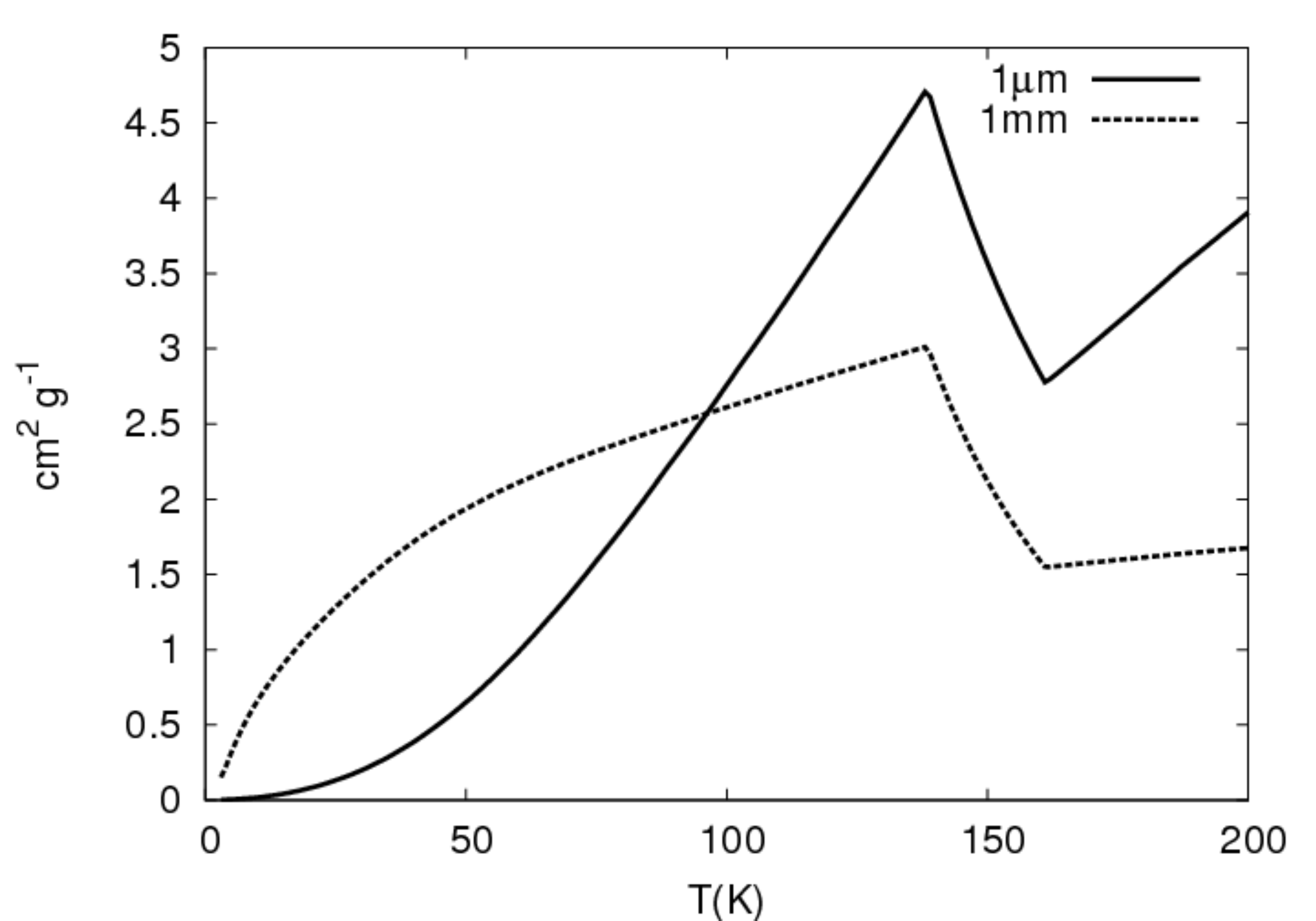}
\caption{Opacities from the 1mm and $1\mu$m 
tables.  For the temperatures in the spiral in these simulations, the 1 mm case is much more opaque.  Only after clump formation do temperatures rise enough to make the 1mm opacities less opaque than the $1\mu$m. }
\end{figure}

\begin{figure}
\includegraphics[width=15cm]{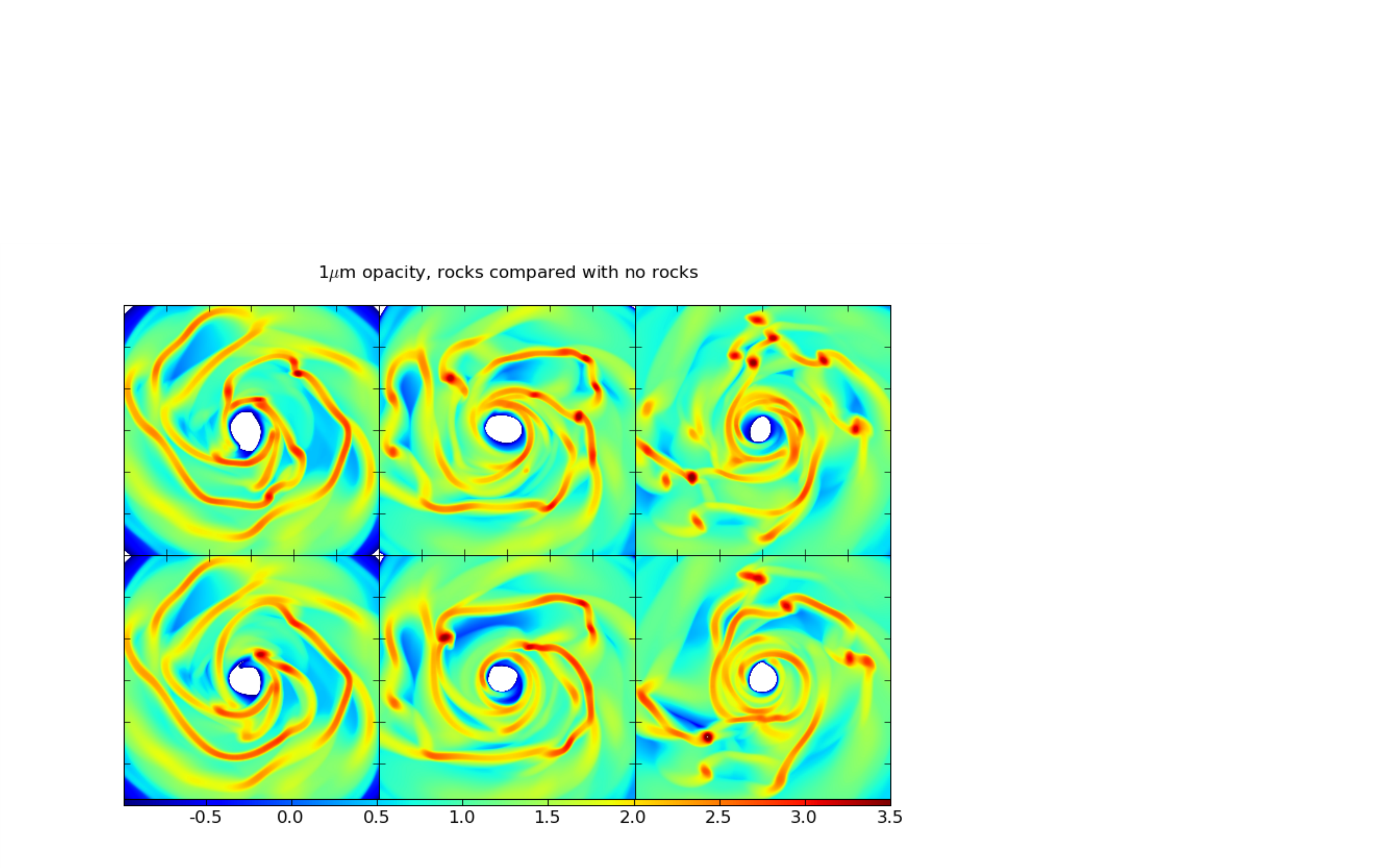}
\caption{
Gas surface density snapshots for SIM1mu (top) and SIM1muNoRocks (bottom). Each square is 300 AU on a side, and the snapshots correspond to 1260, 1410, and 1570 yr going from left to right. The colorbar shows surface density in g cm$^{-2}$.  The evolutions are qualitatively similar between the disks, but differences, e.g., the number of fragments, are apparent.}
\end{figure}

\begin{figure}
\includegraphics[width=15cm]{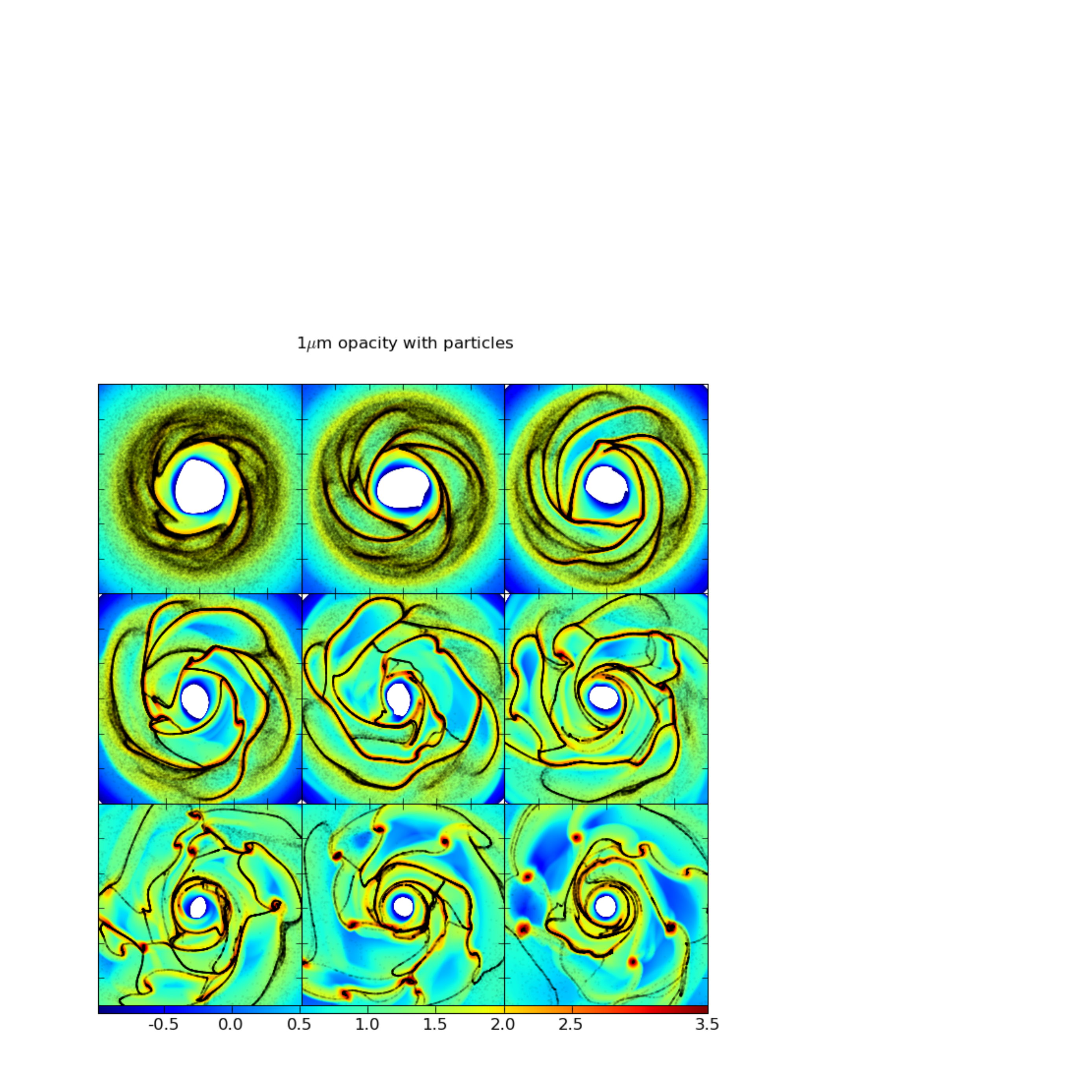}
\caption{
Gas surface density snapshots with rock positions superposed for SIM1mu. Each square is 300 AU on a side, and the snapshots correspond to  620, 780, 940, 1100, 1260, 1410, 1570, 1720, and 1870 yr going from left to right and top to bottom. The colorbar shows the logarithmic surface density in  g cm$^{-2}$. }
\end{figure}

\begin{figure}
\includegraphics[width=8cm]{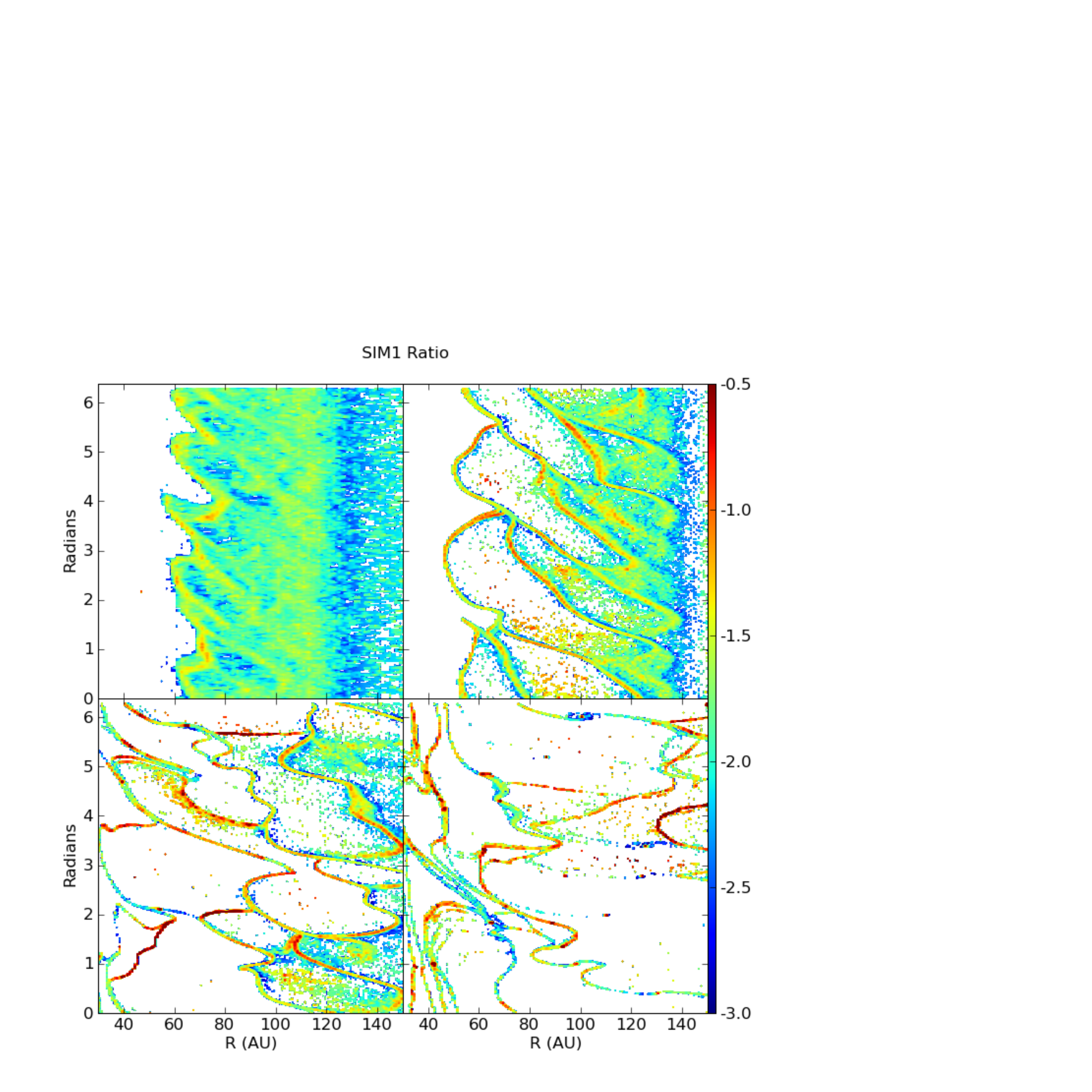}\includegraphics[width=8cm]{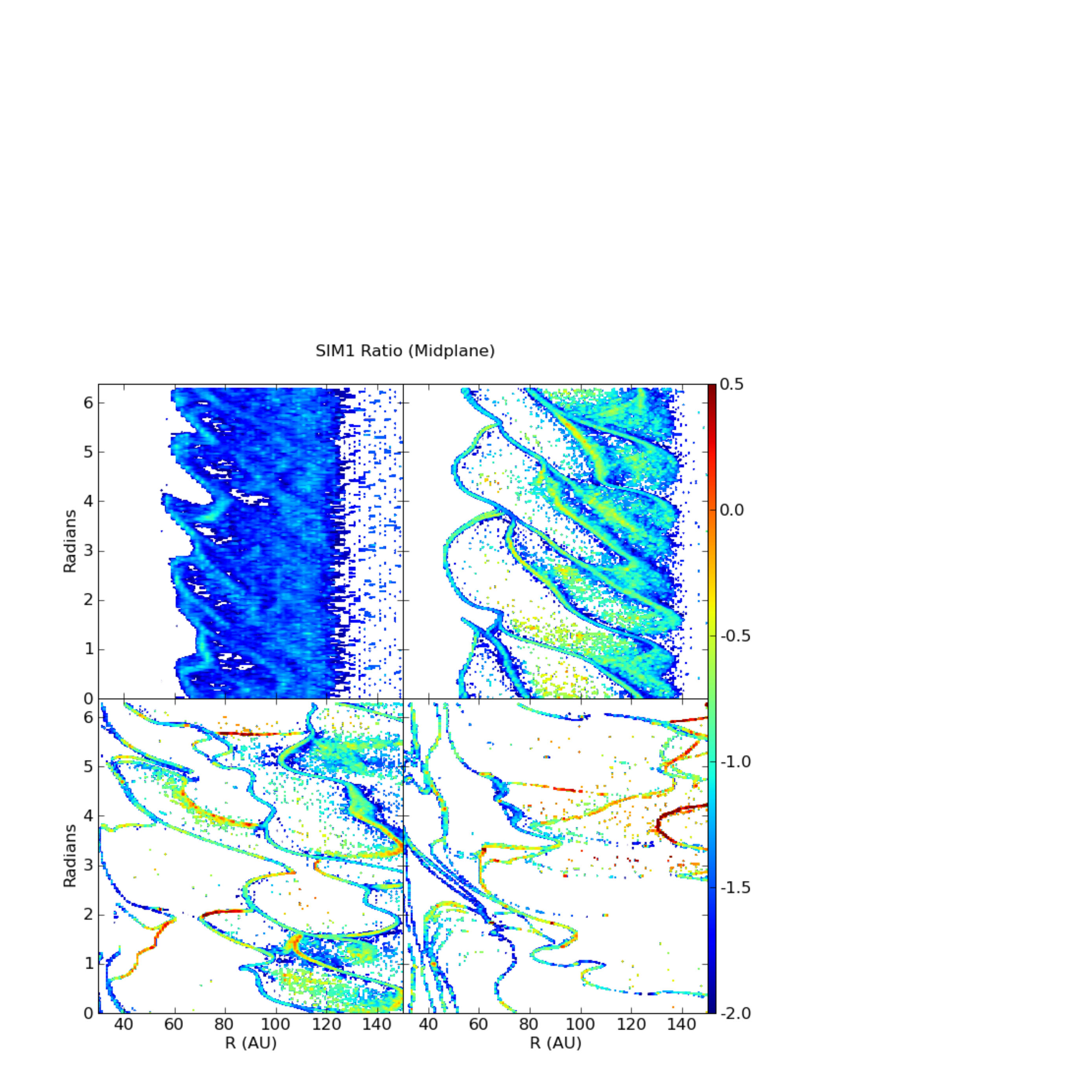}
\caption{The logarithm of the rock-to-gas ratio in cylindrical coordinates for several snapshots 
of SIM1mu.   The spiral arms show large surface density enrichment, while the interarm 
regions show significant solid depletion.  The snapshots are for 620, 940, 1260, and 1870 yr, from left to right and top to bottom. The left panel shows the ratio of the full surface densities, while the right panel shows the ratio only for the midplane cells.}
\end{figure}

\clearpage

\begin{figure}
\includegraphics[height=8cm,angle=-90]{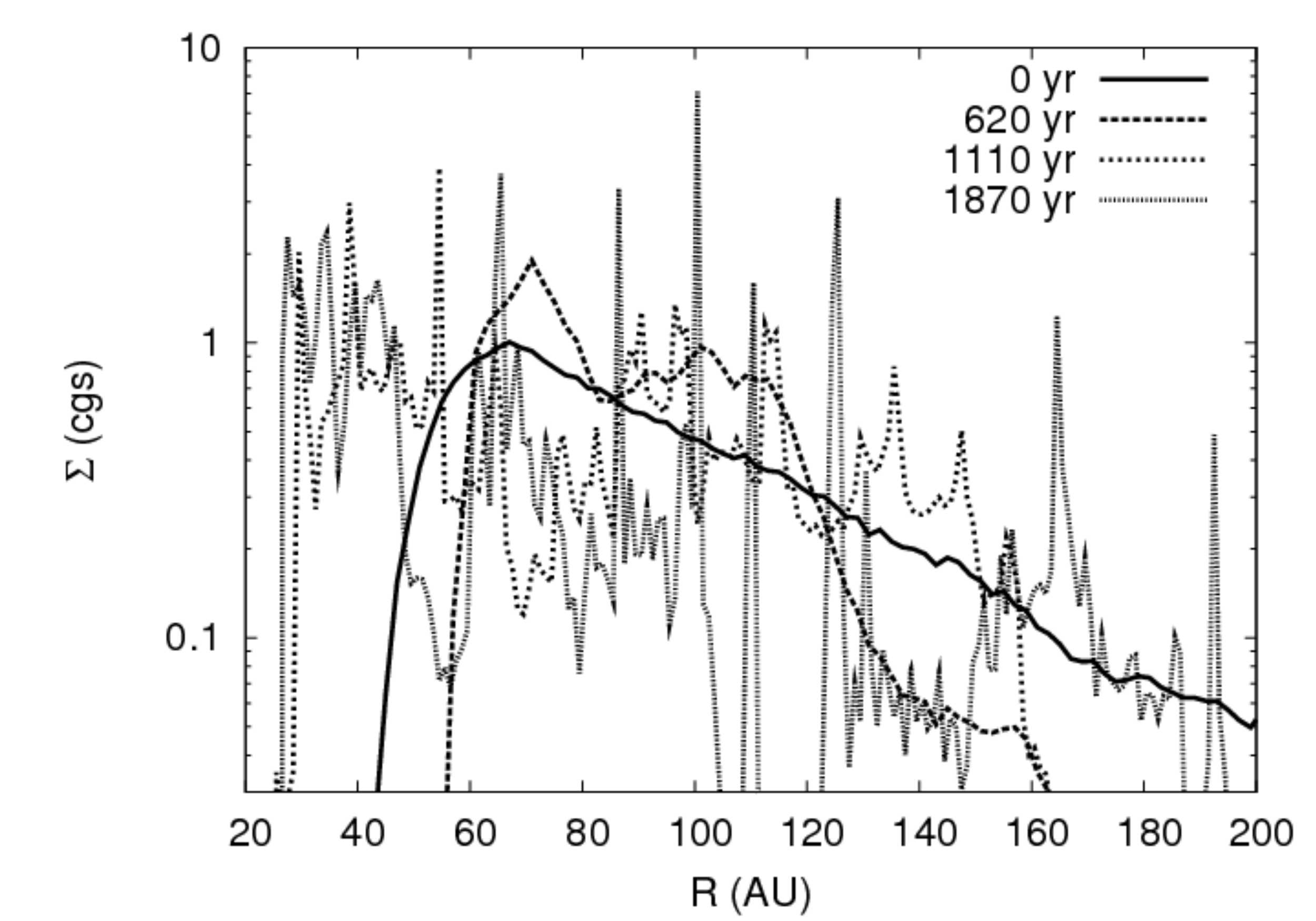}\includegraphics[height=8cm,angle=-90]{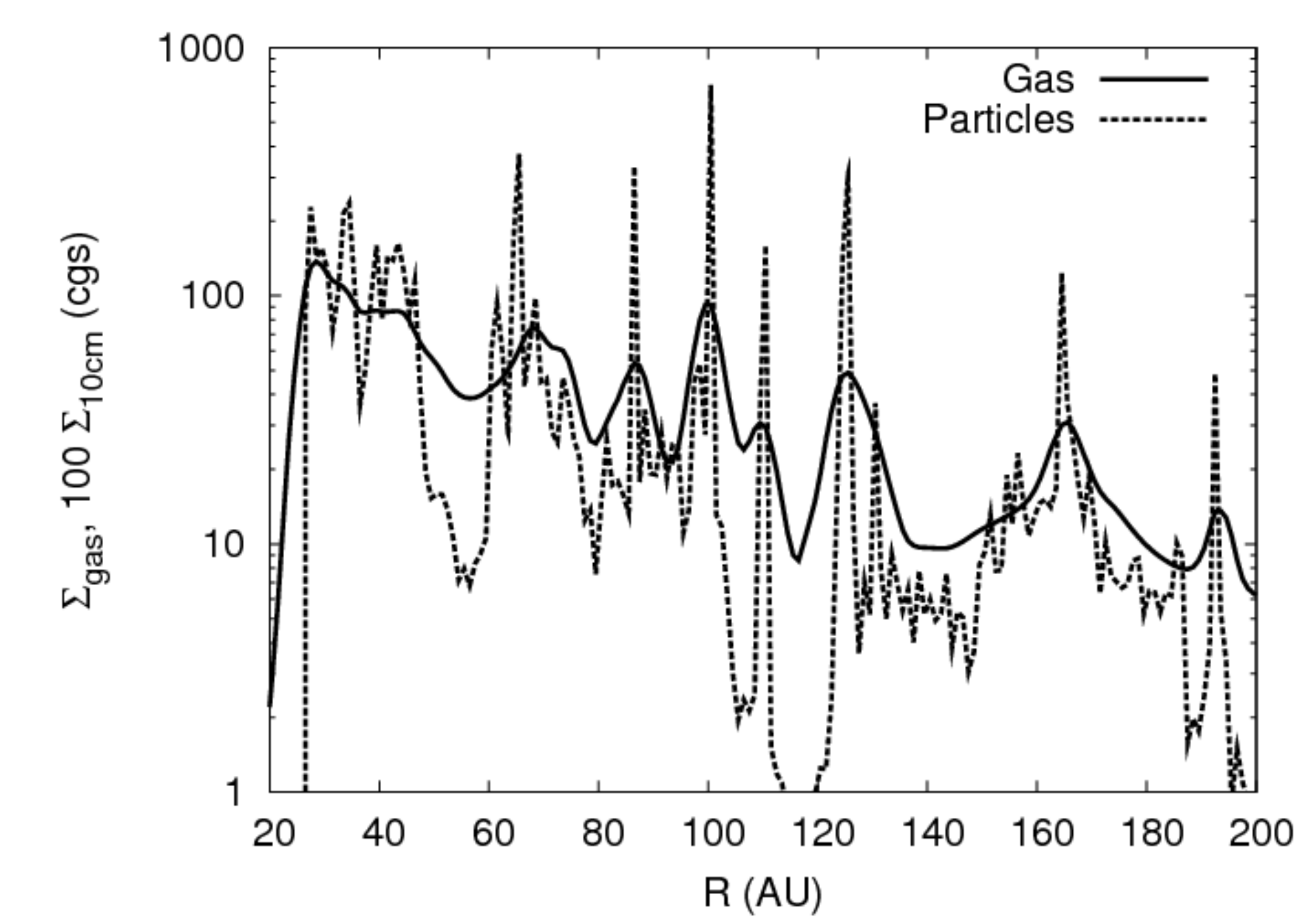}
\caption{Left: Azimuthally averaged surface density profiles of the rocks in SIM1mu.  As spiral arms develop, the radial surface density profiles show strong radial concentrations of rocks and redistribution of material outward as well as inward. Right: The azimuthally averaged gas surface density shown with the rock surface density, multiplied by 100, at 1870 yr in SIM1mu.  Gas surface density peaks show corresponding rock density enhancements.  }
\end{figure}

\begin{figure}
\includegraphics[width=15cm]{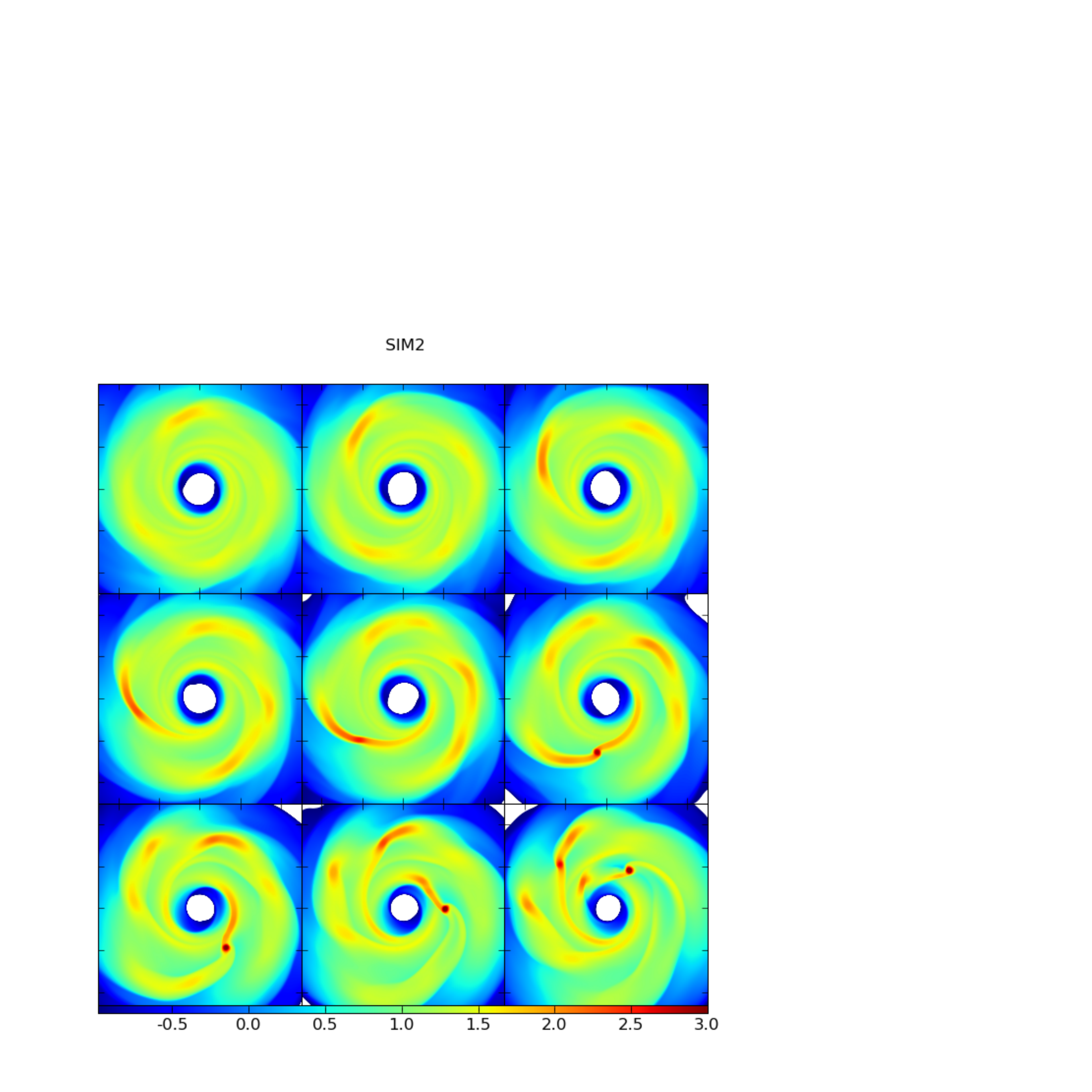}
\caption{
Gas surface density snapshots for SIM2. Each square is 500 AU on a side, and the snapshots correspond to  320, 450,  640, 800, 960, 1130, 1300, 1450, and 1620  yr going from left to right and top to bottom. The colorbar shows the logarithmic surface density in g cm$^{-2}$.  }
\end{figure}

\clearpage

\begin{figure}[ht*]
\includegraphics[width=15cm]{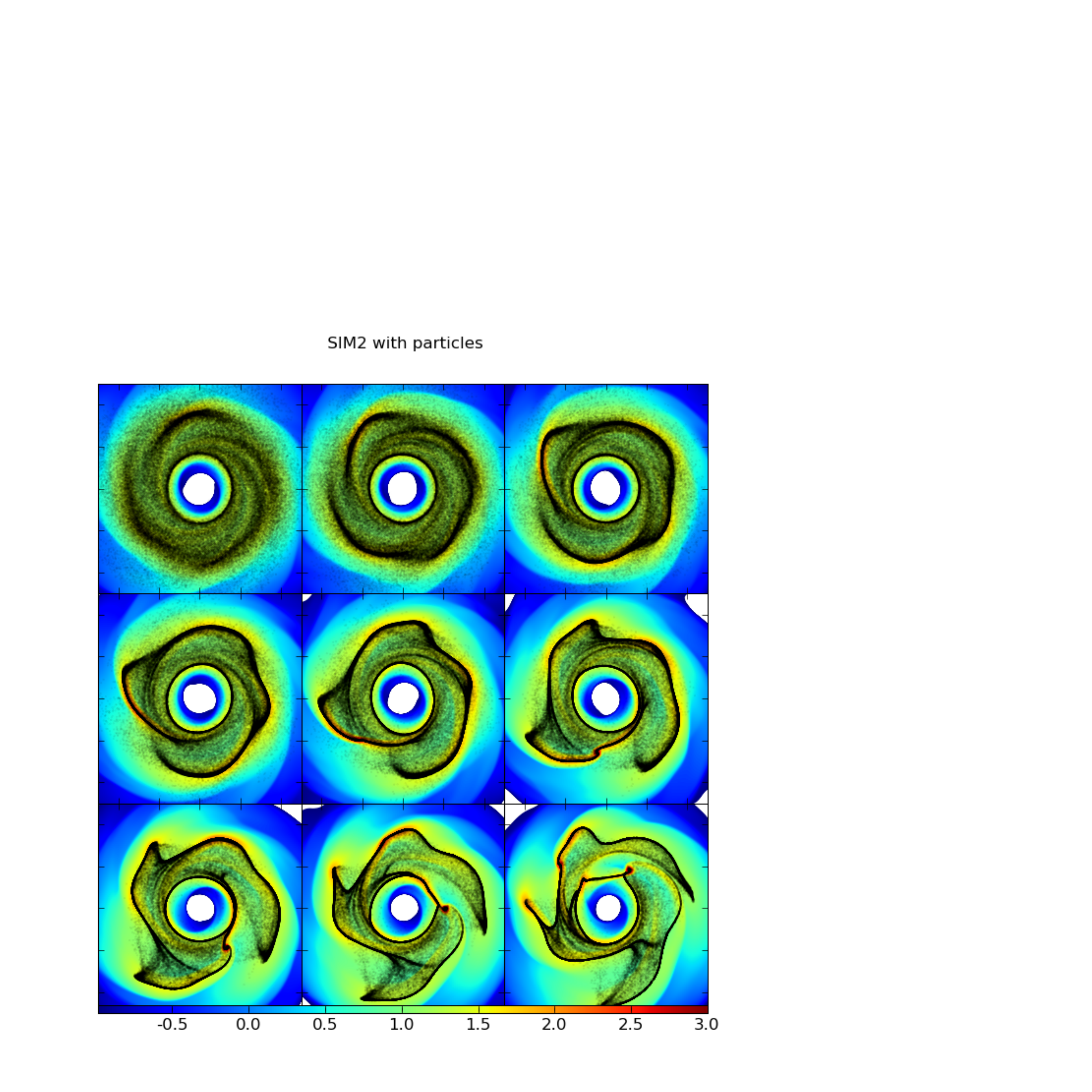}
\caption{The same as in Figure 13, but with the particle distribution superimposed.}
\end{figure}

\begin{figure}[ht*]
\includegraphics[width=8cm]{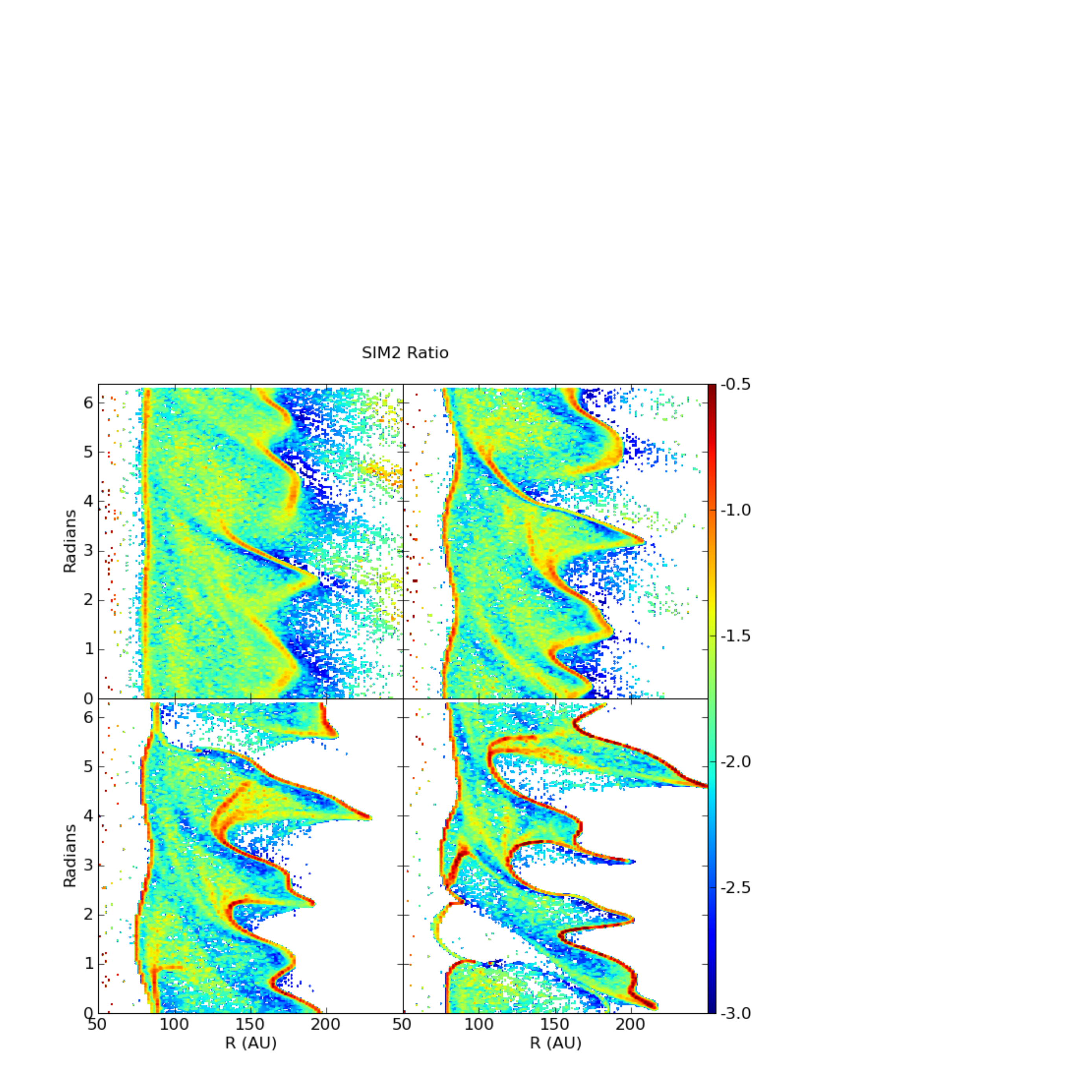}\includegraphics[width=8cm]{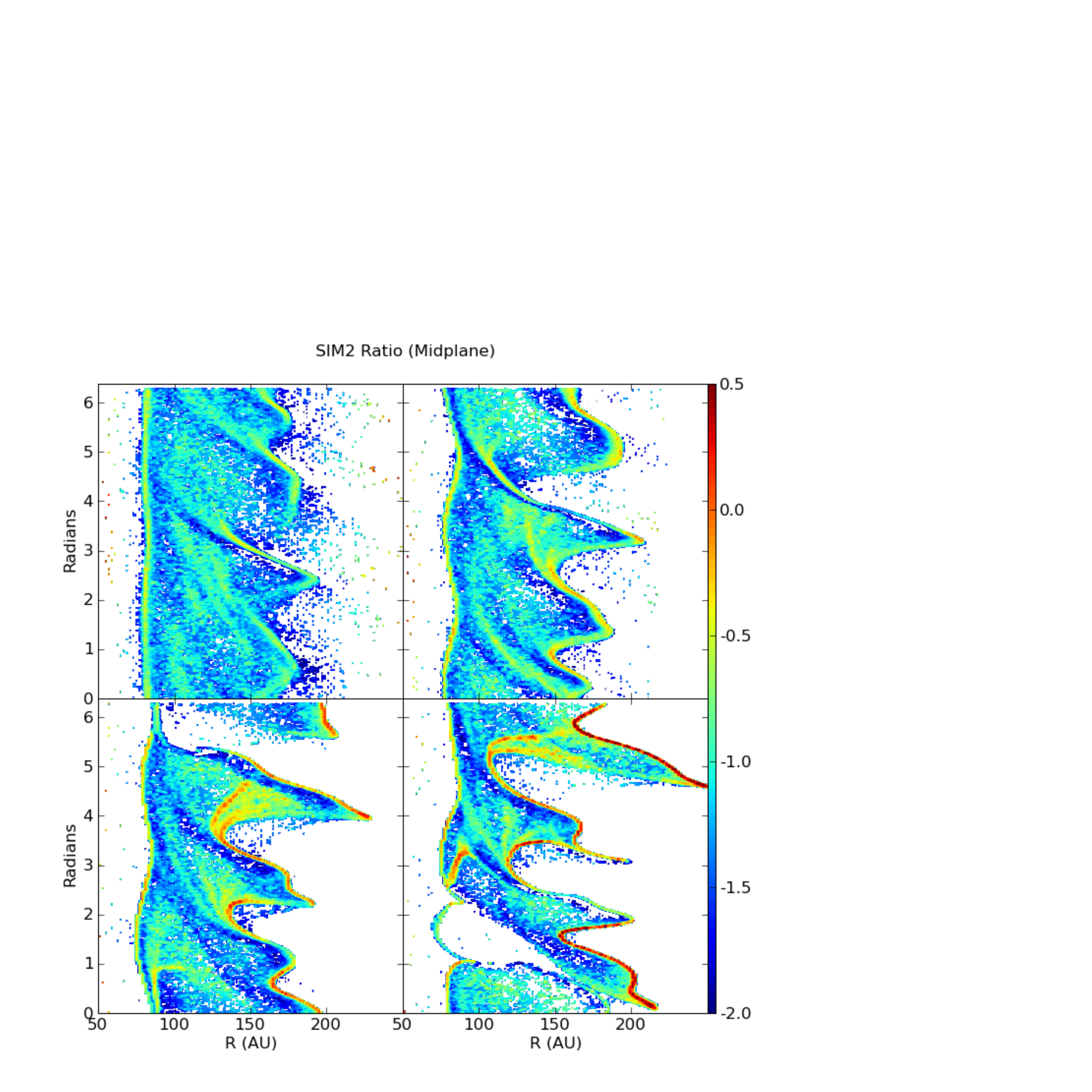}
\caption{The logarithm of the rock-to-gas ratio in cylindrical coordinates for the SIM2
snapshots 640, 9600, 1300, and 1620 yr. The left panel shows the ratio of the full surface densities, while the right panel is only for the midplane cells.}
\end{figure}

\begin{figure}[ht*]
\includegraphics[height=8cm,angle=-90]{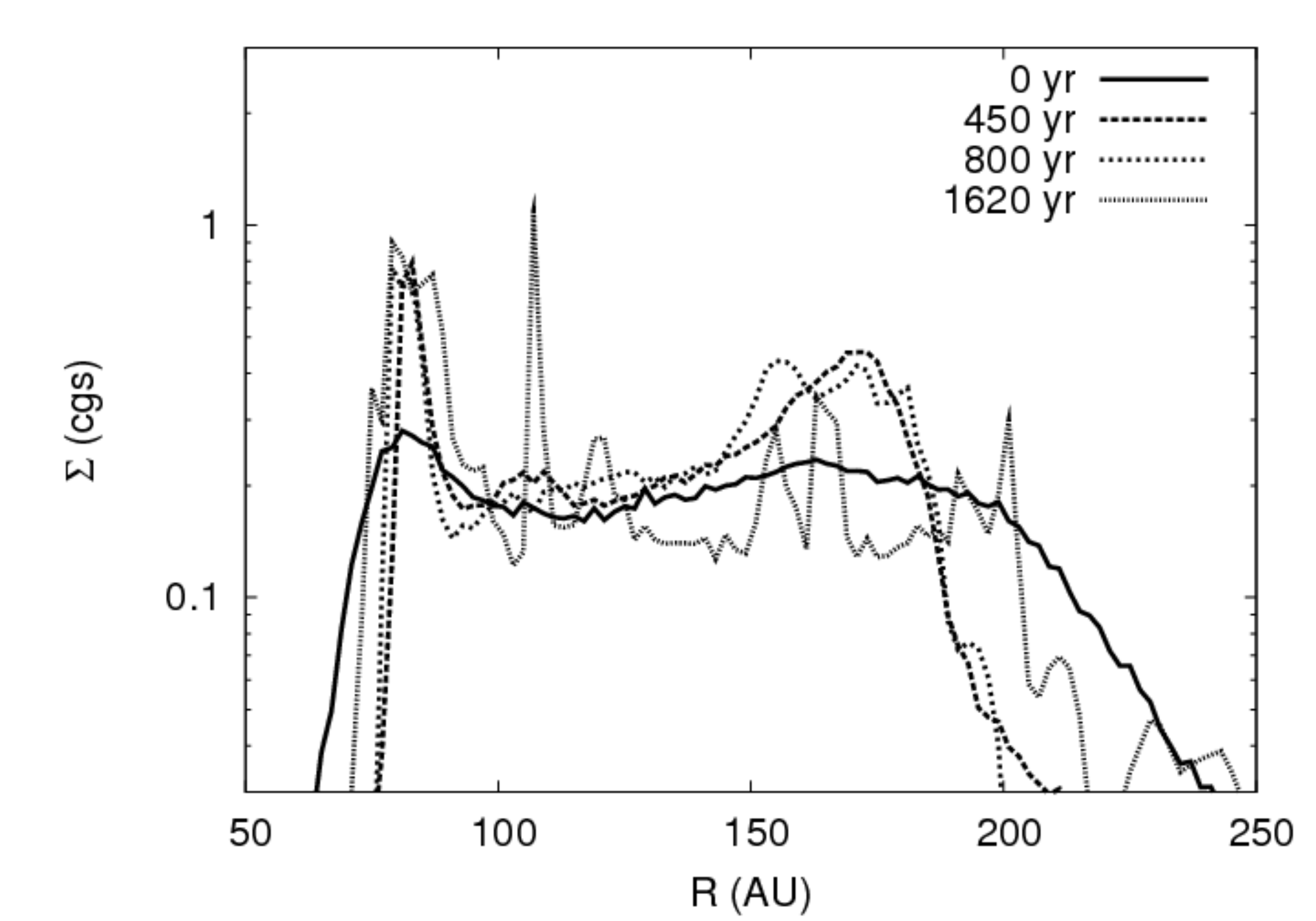}\includegraphics[height=8cm,angle=-90]{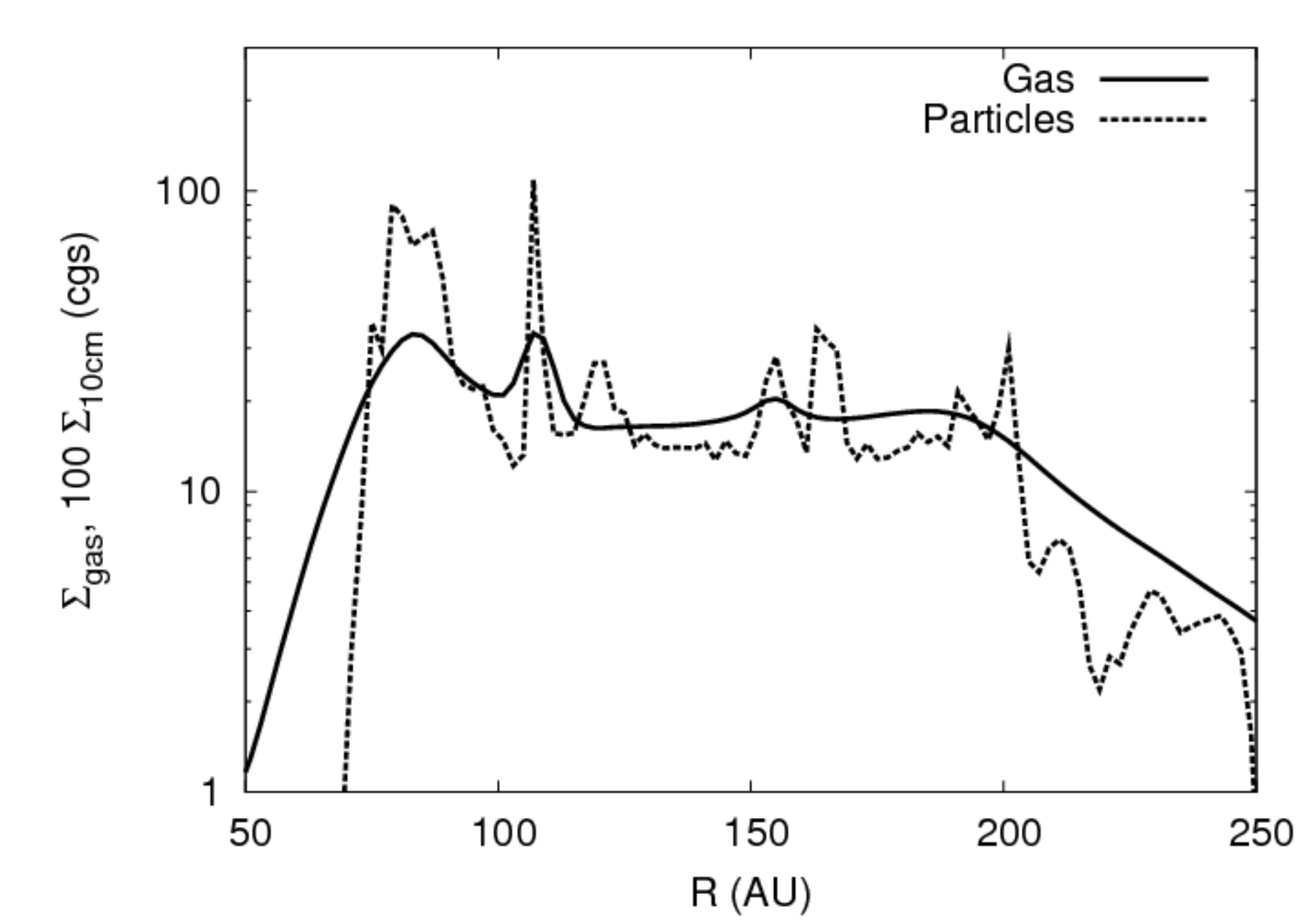}
\caption{Left: Azimuthally averaged surface density profiles of the rocks in SIM2.  As in SIM1mu, spiral arms concentrate and redistribute the rocks. 
Right: The azimuthally averaged gas surface density shown along with the rock surface density, multiplied by 100, at 1620 yr in SIM2.  Gas surface density peaks show corresponding rock density enhancements. }
\end{figure}

\begin{figure}[ht*]
\includegraphics[width=15cm]{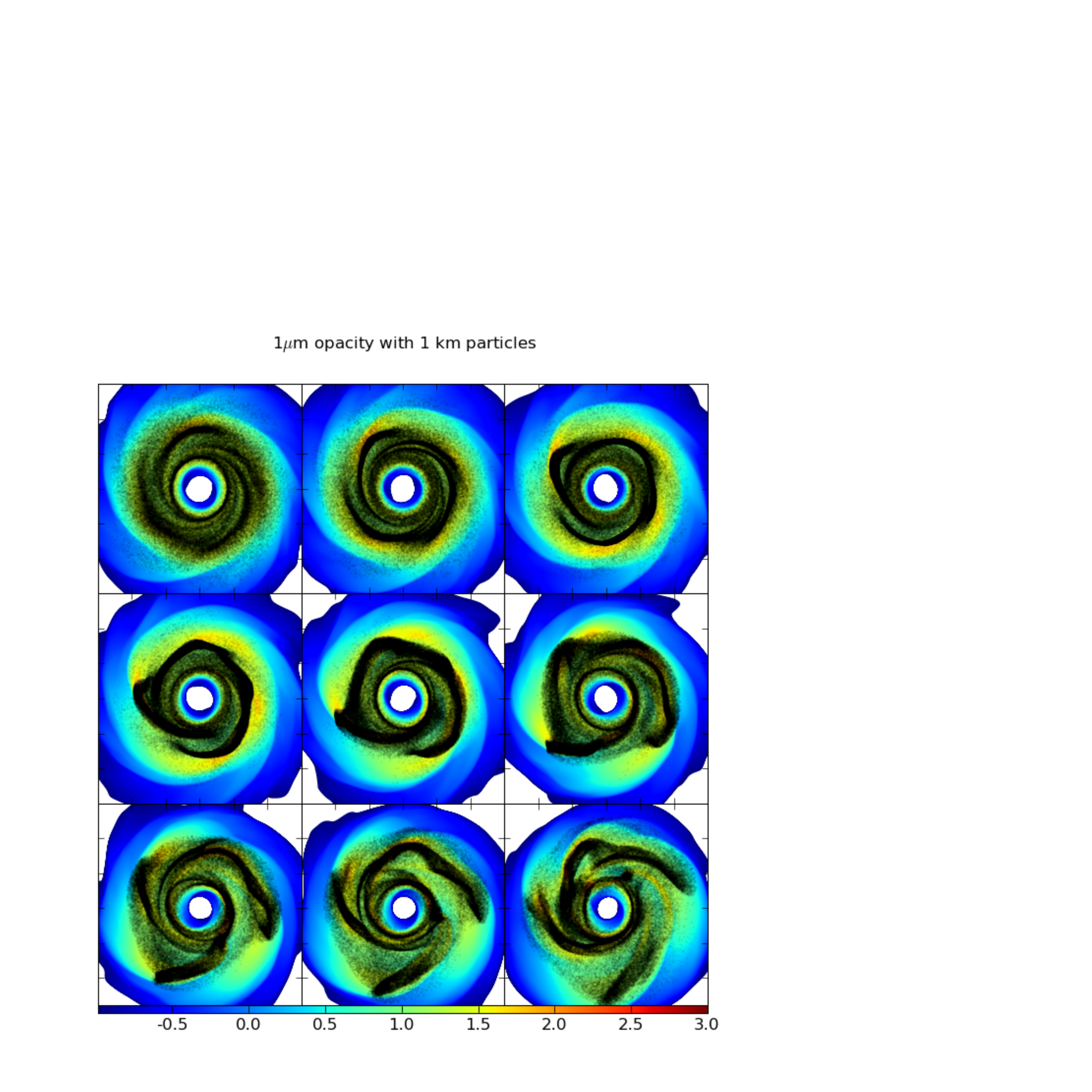}
\caption{Similar as Figure 14, but  for SIM2km.}
\end{figure}

\begin{figure}[ht*]
\includegraphics[width=12cm]{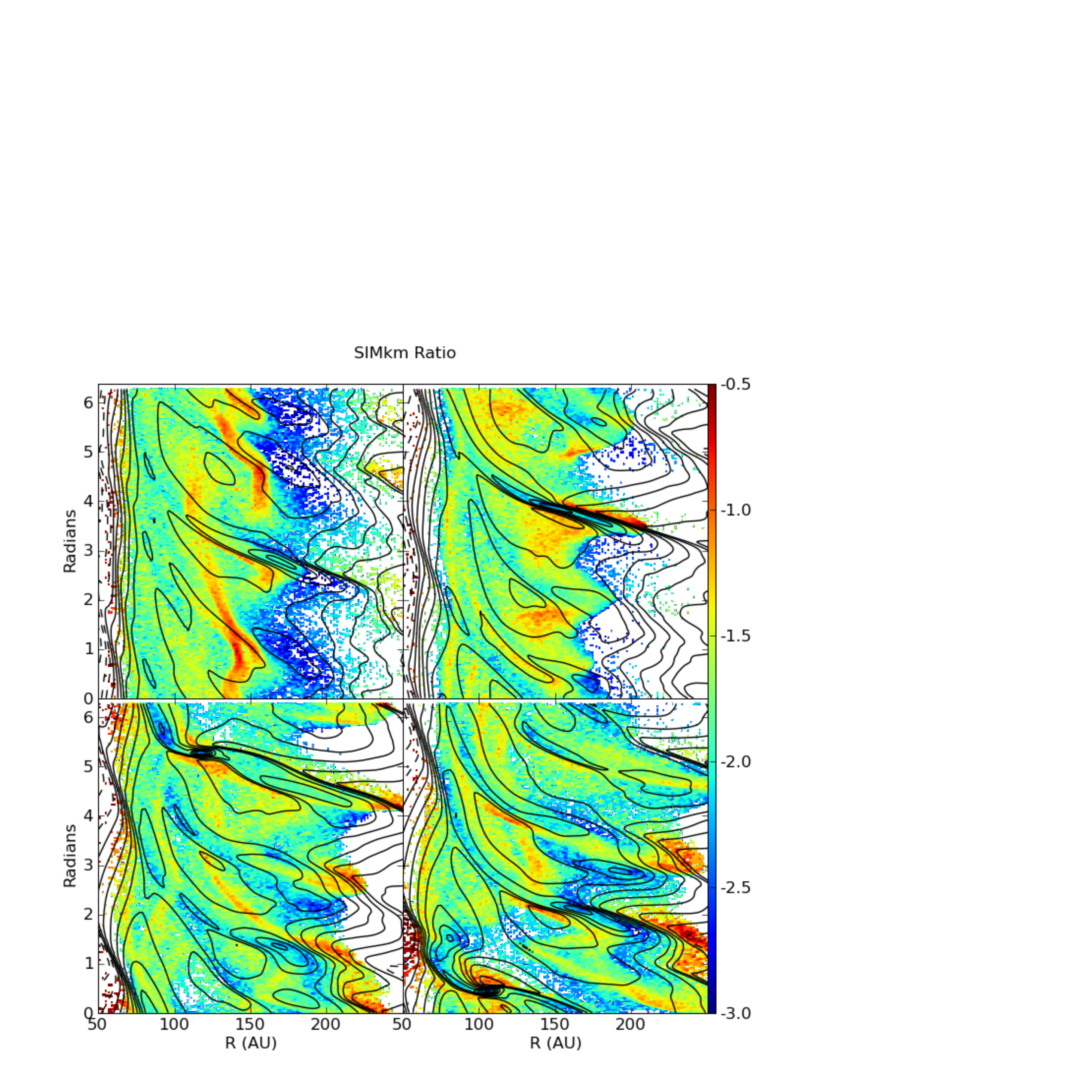}
\caption{The logarithm of the rock-to-gas ratio in cylindrical coordinates for SIM2km.  The snapshots are at similar times as shown in Figure 15.
Only the ratio of the full surface densities is shown.  The contours represent the surface density of the gas, demonstrating that the regions with enhanced solid-to-gas ratios are outside the gaseous spiral arms. In addition, although the clump itself is slightly depleted in solids, the regions immediately around the clump do have super solids-to-gas ratios.}
\end{figure}

\begin{figure}[ht*]
\includegraphics[width=9cm]{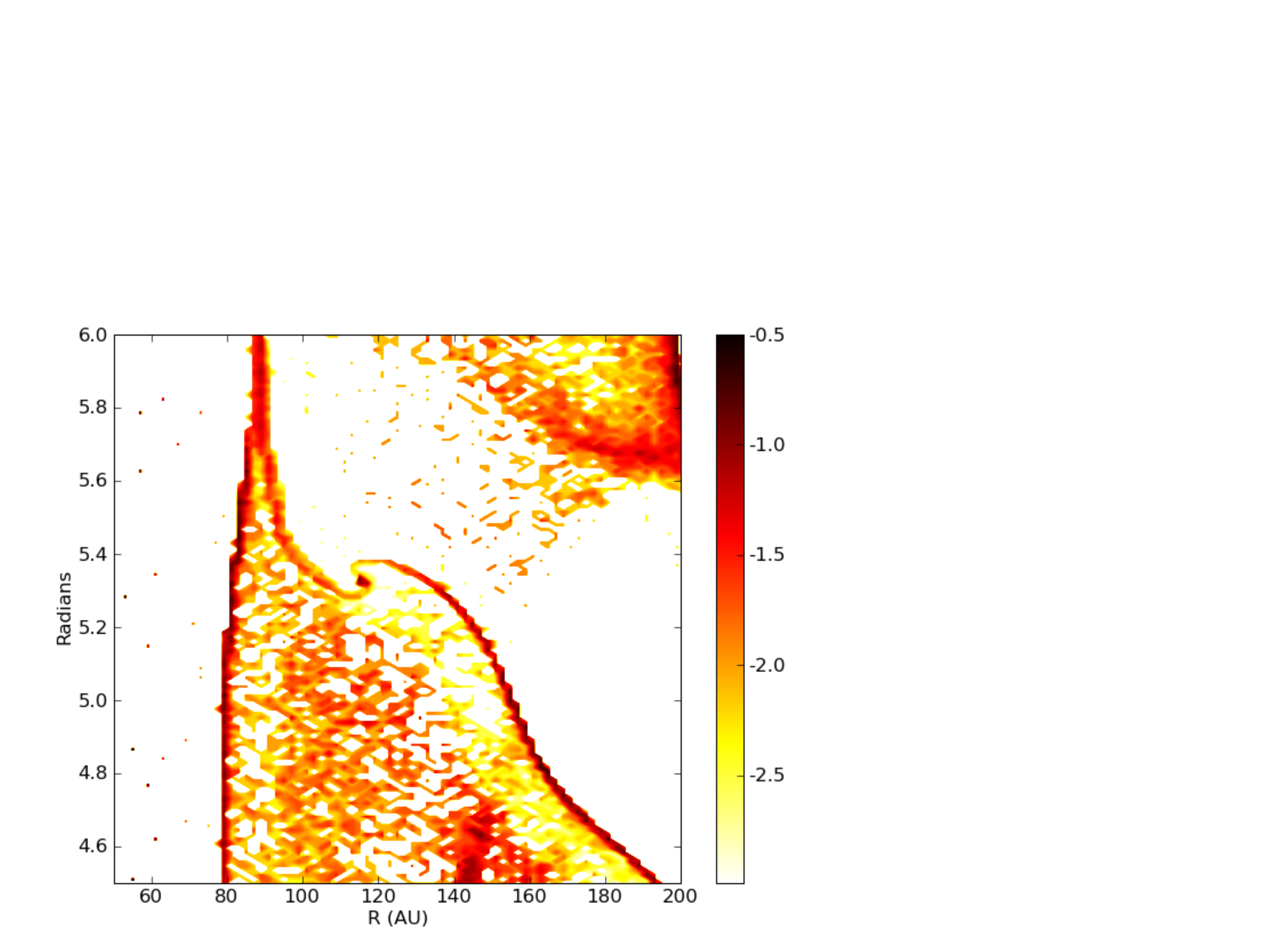}\includegraphics[width=9cm]{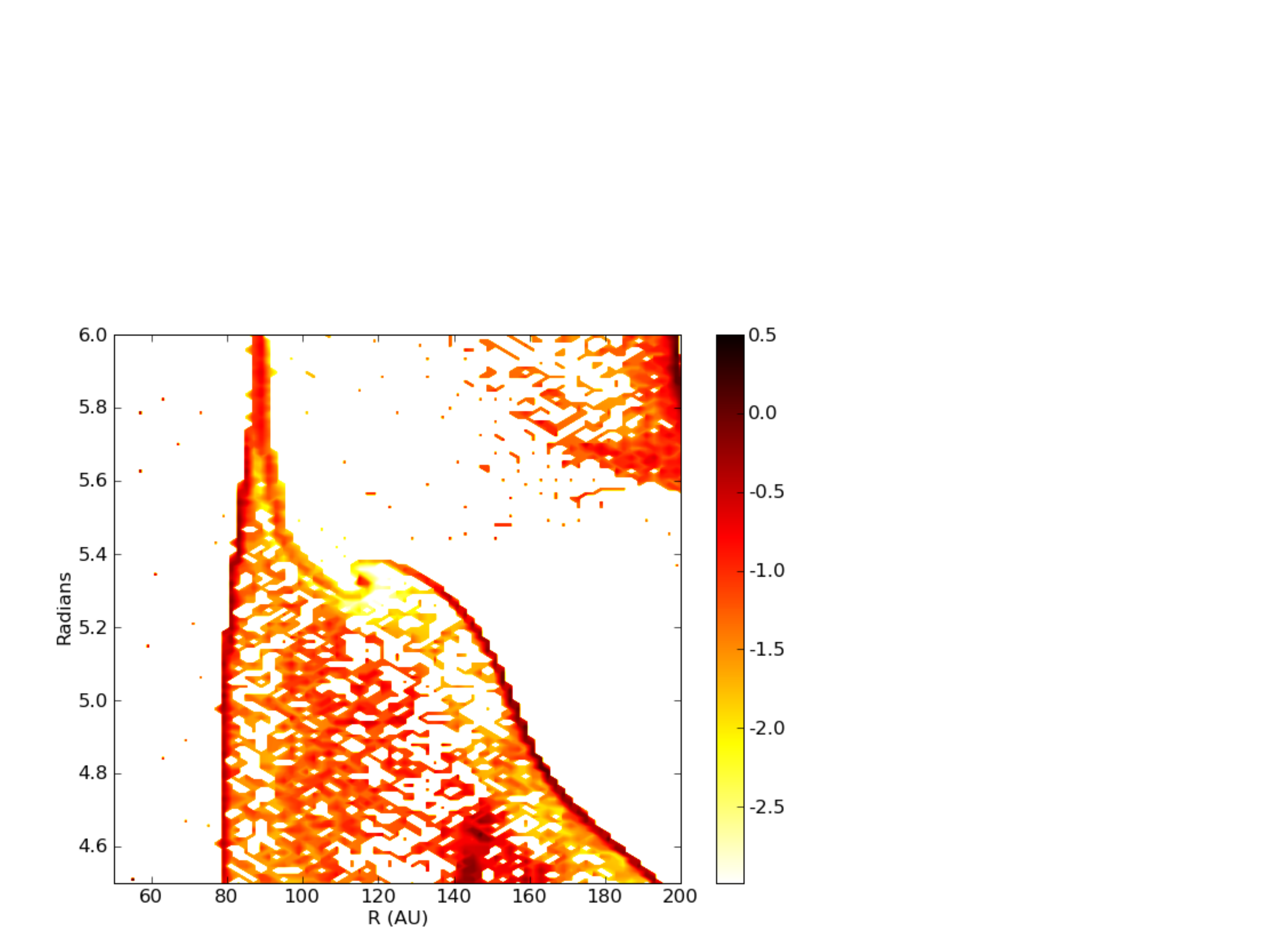}
\caption{Similar to Figure 15, but 
for a region zoomed in around the clump in SIM2.  The left panel shows 
the logarithm of the rock-to-gas ratio for the entire surface density, while the right panel is just for the midplane cells. }
\end{figure}

\clearpage

\begin{figure}[ht*]
\includegraphics[width=6cm,angle=-90]{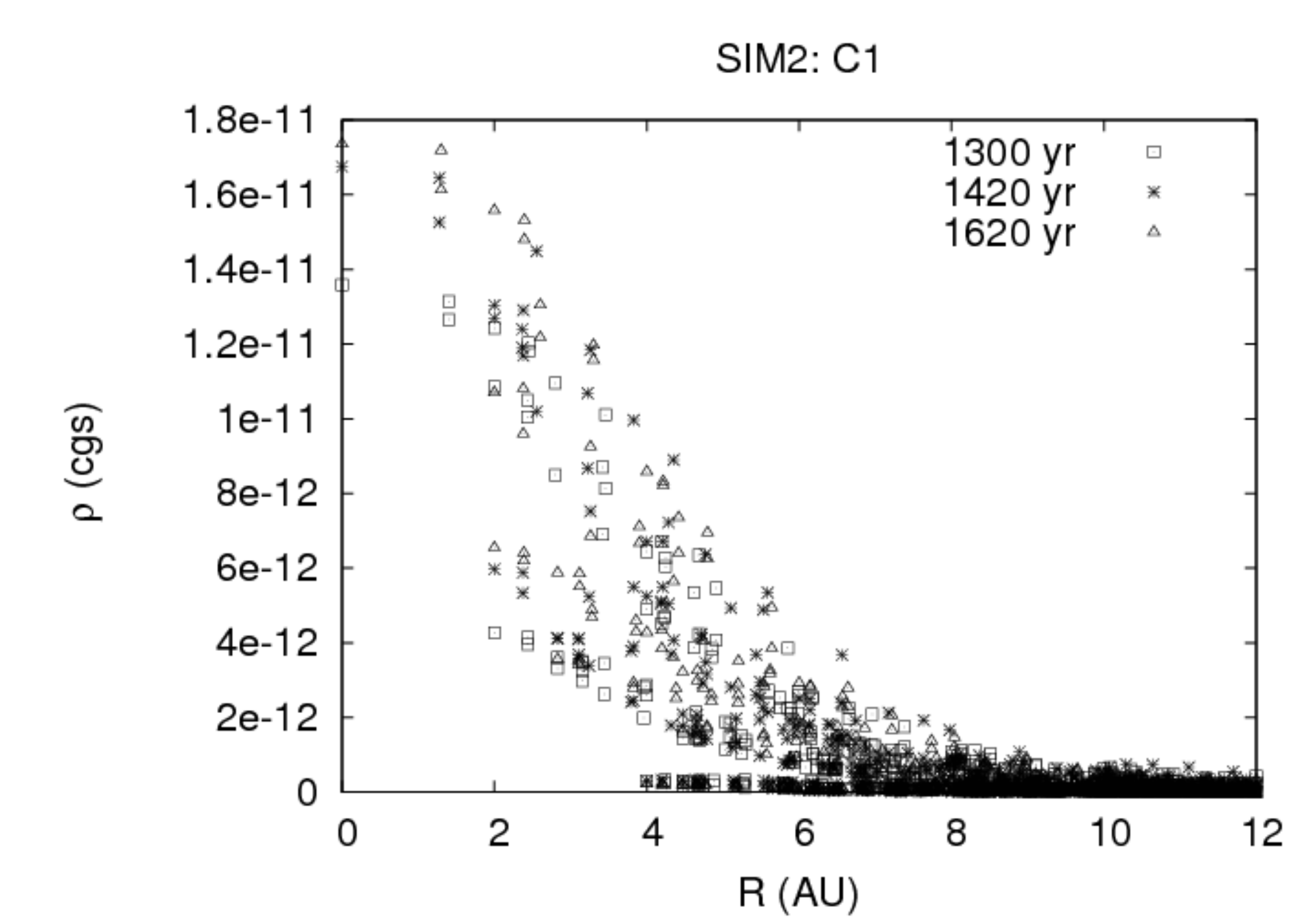}\includegraphics[width=6cm,angle=-90]{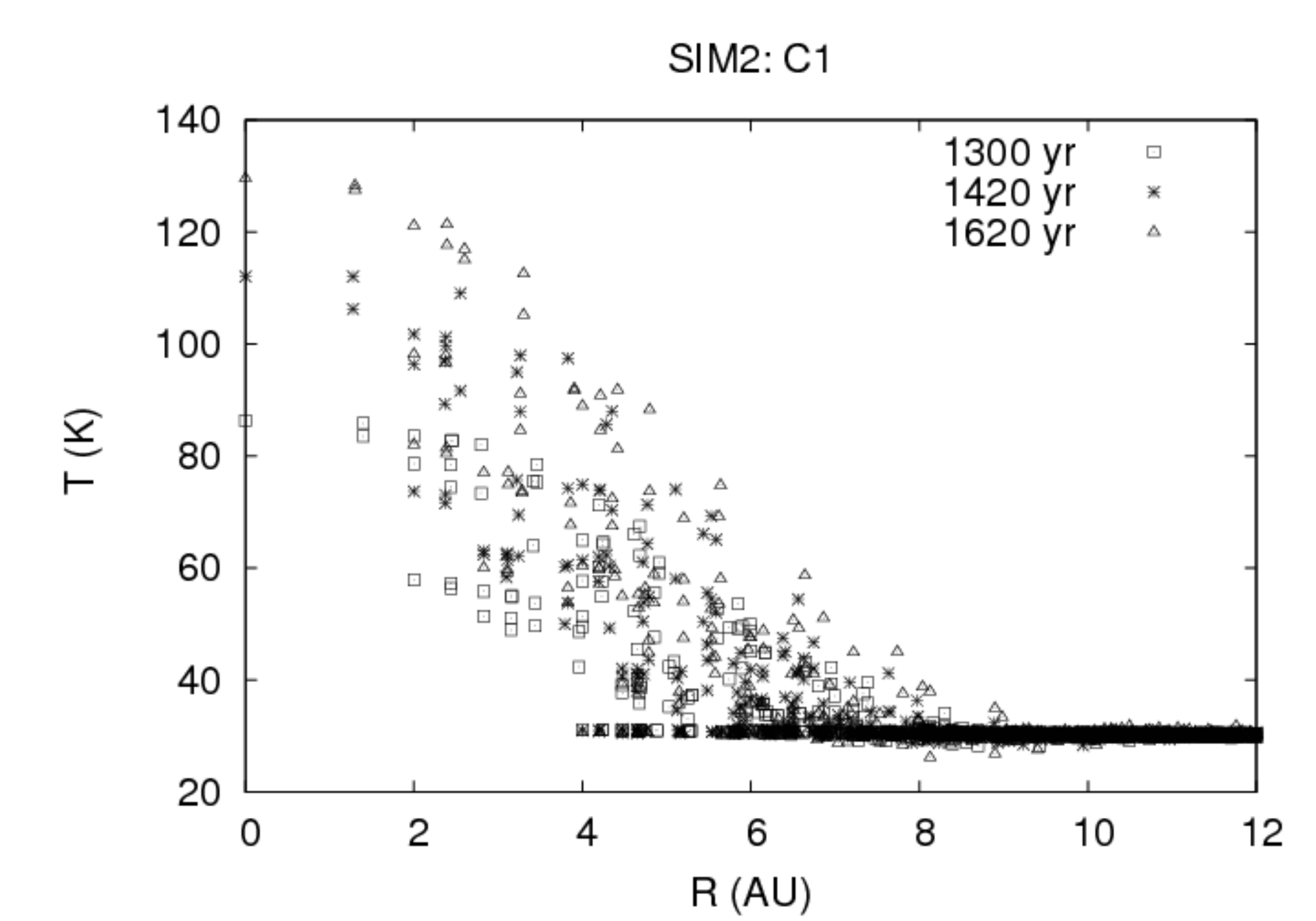}
\caption{Cell values for the gas density and temperature as a function of distance from the density maximum in the SIM2:C1 clump. Several snapshots are shown.}
\end{figure}

\begin{figure}[ht*]
\includegraphics[width=9cm]{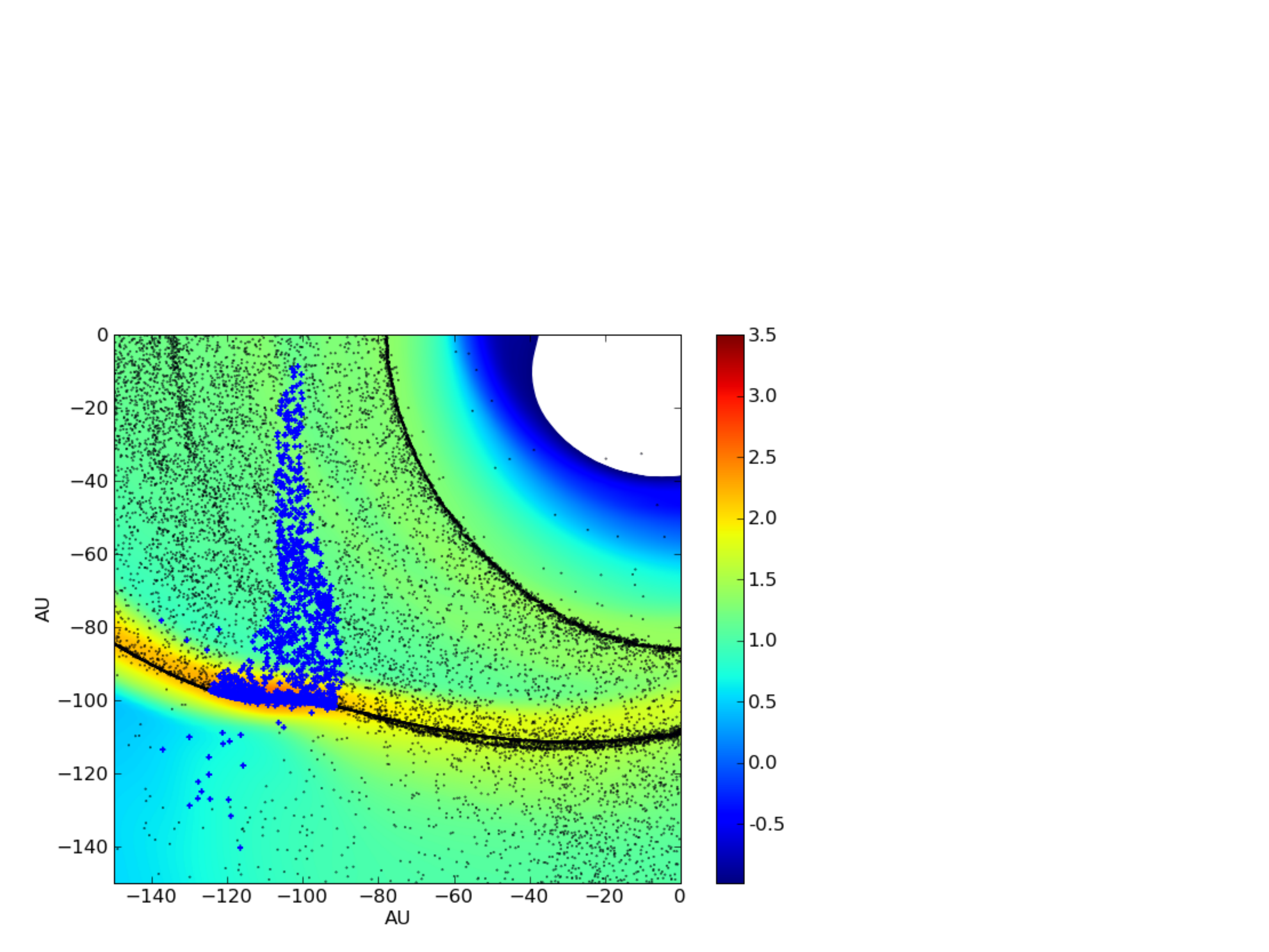}\includegraphics[width=9cm]{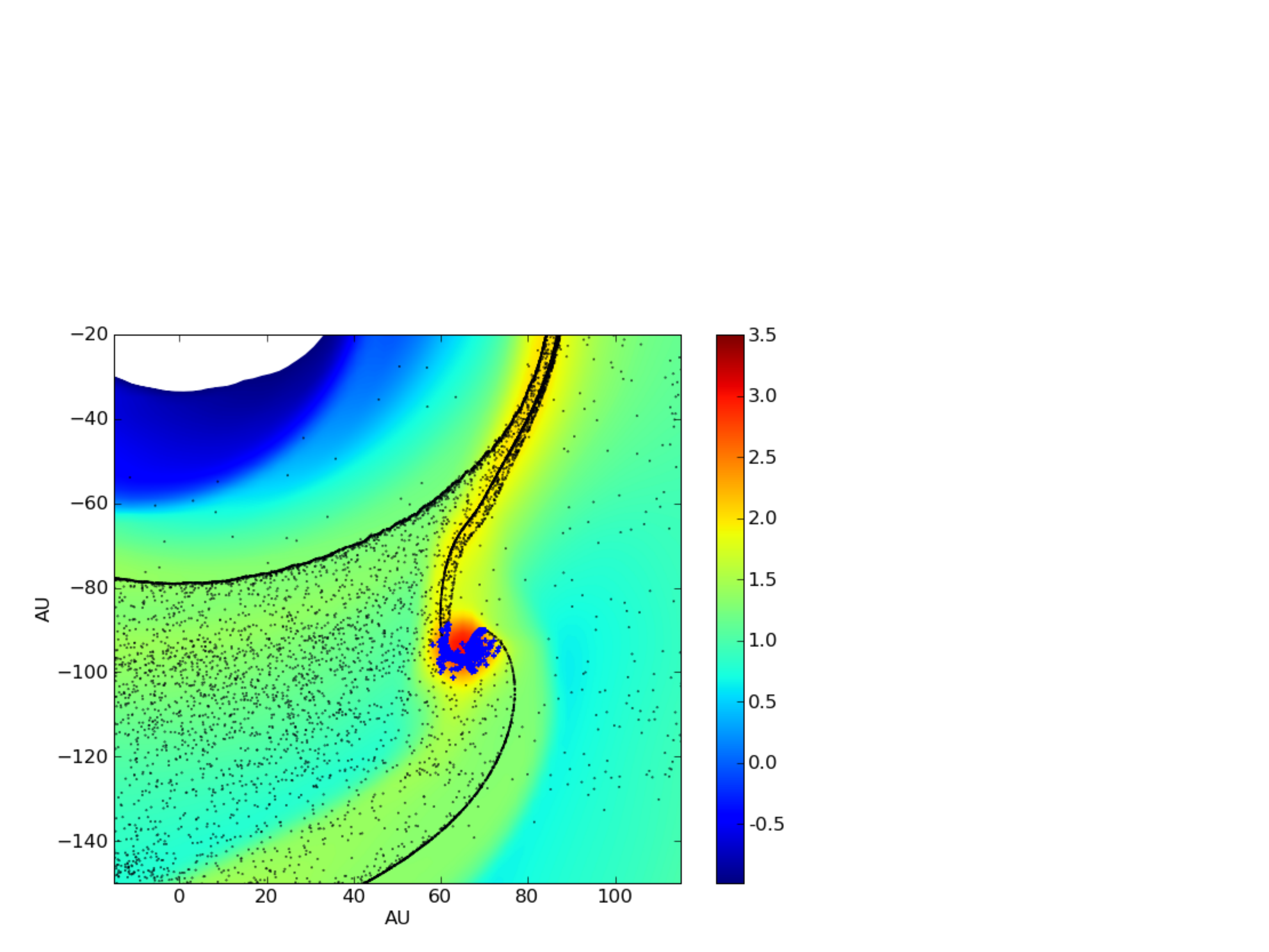}
\caption{The spiral arm in SIM2 just before fragmentation at 960 yr (left) and after fragmentation at the 1300 yr snapshot (right).  Gas density is shown by the colorbar, while rock particles are given by black dots.  The rocks that are considered to be part of the clump are shown with large, blue crosses.  The same particles are shown in the 960 yr snapshot.  Most of the mass comes from a 20-30 AU section of the spiral arm.}
\end{figure}

\begin{figure}[ht*]
\includegraphics[width=15cm]{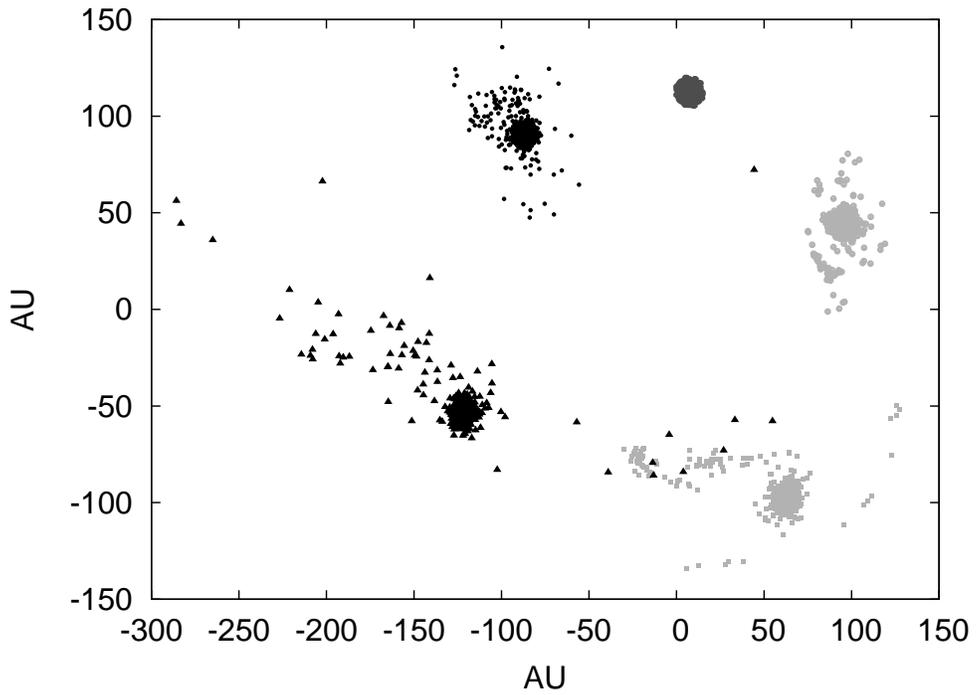}
\caption{Solids associated with SIM2km:C1 at 1720 yr, and then traced back to an early stage of its orbit (light gray) and to later stages of the clump's orbit (black). Particles coming into the clump are not easily trapped, and can be scattered far from the planet's Hill sphere. The earliest snapshot shown is 1290 yr, and the latest is 2190 yr.}
\end{figure}

\begin{figure}[ht*]
\includegraphics[width=6cm,angle=-90]{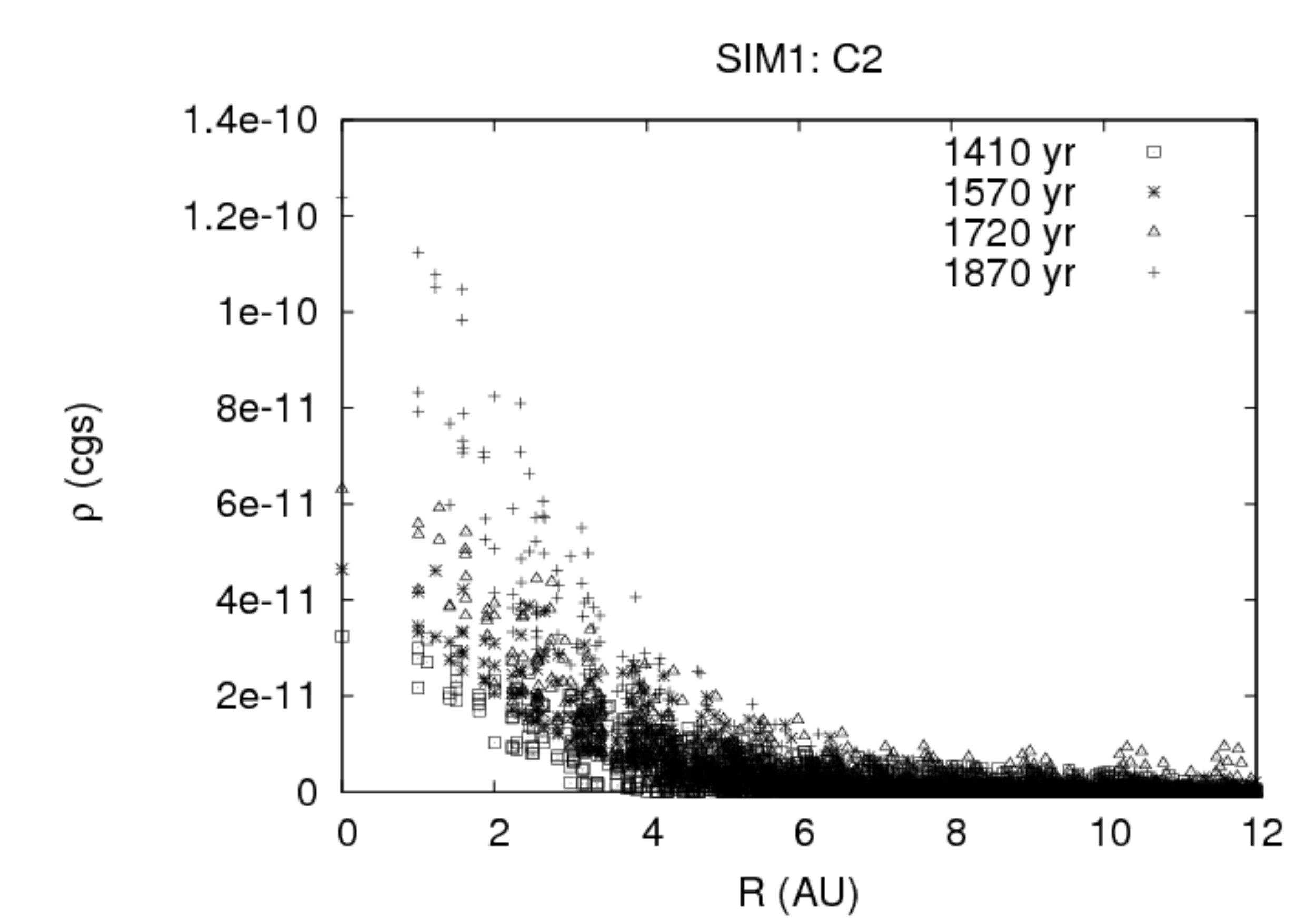}\includegraphics[width=6cm,angle=-90]{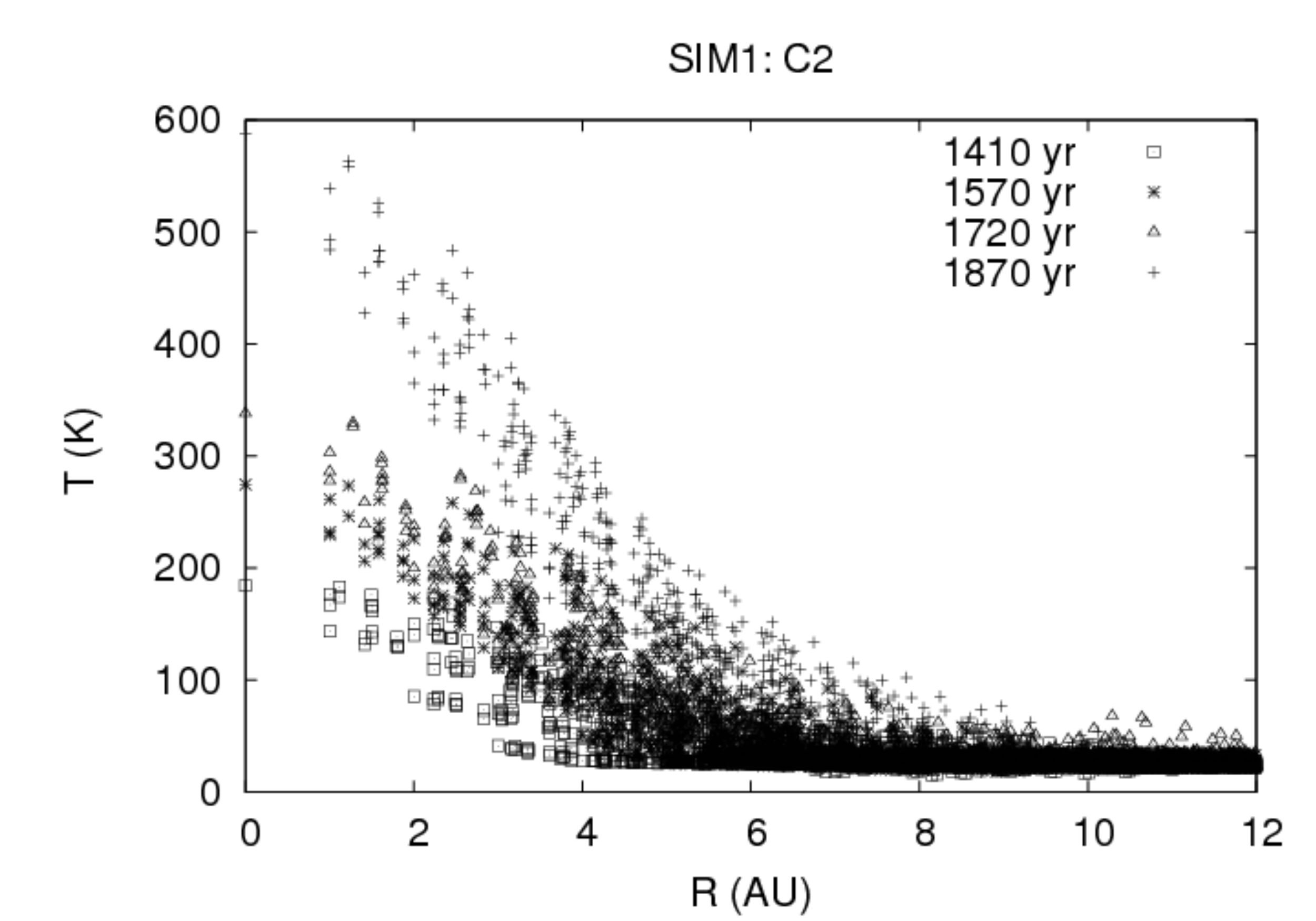}
\caption{Cell values for the gas density and temperature as a function of distance from the density maximum in the SIM1mu:C2 clump. Several snapshots are shown.}
\end{figure}

\clearpage

\begin{figure}
\includegraphics[width=12cm]{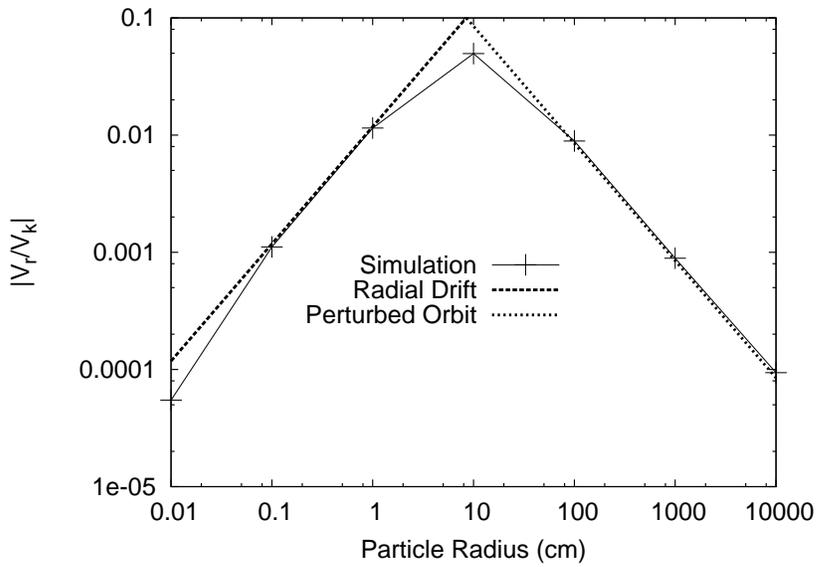}
\caption{Actual versus analytic asymptotic radial drift speeds for the solids.  At very small particle sizes, the stopping times become comparable to the hydrodynamics time step, which causes more solid-gas coupling than expected.  For the regime explored in the simulations presented in this paper, the algorithm is acceptable.}
\end{figure}

\clearpage

\begin{table}
\caption{Gas and rock mass for the clump that forms in SIM2, as well as one comparison with SIM2km.  The disk radius $r$ and azimuthal angle $\phi$ of the clump are given for each data set, and can be compared with Figure 13.  Three different density thresholds are shown, where only mass above the threshold is considered.  Full represents the entire mass of the clump, here for all gas that has $\rho>9\times 10^{-13}$ g cm$^{-3}$ and for $T>34$ K.  Half refers to all mass that is above half of the peak density in the clump, $\rho_{\rm max}$, and $\gtrsim 90\%$ is for only mass that is greater than 90\% of the peak.  This last threshold isolates the core conditions of the clump. The factor $f_{\rm RtoG}$ gives the ratio of rock mass to gas mass, and $f_{\rm enr}\equiv(f_{\rm RtoG}+0.01)/0.02$ gives the total solids enrichment relative to the average nebula's value. For our conversion, we used $1 M_J=320 M_{\oplus}$. 
We remind the reader that when we write ``rocks,'' we are referring to a mixture of silicates and ices. }
\begin{tabular}{ l l l l  l }\hline

SIM2:C1 & $\rm Time=1300$ yr & $r\sim114$ AU & $\phi\sim 305^{\rm \circ}$ & $\rho_{\rm max}=1.4\times 10^{-11}$ g cm$^{-2}$\\\hline
Density Threshold &  Gas ($M_J$) & Rocks ($M_{\oplus}$) & $f_{\rm RtoG}$ & $f_{\rm enr}$\\\hline
Full & 7.1 & 35 & 0.015 & 1.3 \\
Half&  2.9 & 32 & 0.034 & 2.2 \\
$\gtrsim 90\%$ &  1.0 & 30 & 0.094 & 5.2 \\\hline

SIM2:C1 & $\rm Time=1450$ yr & $r\sim104$ AU & $\phi\sim 359^{\rm \circ}$ & $\rho_{\rm max}=1.7\times 10^{-11}$ g cm$^{-2}$\\\hline
Density Threshold &  Gas ($M_J$) & Rocks ($M_{\oplus}$) & $f_{\rm RtoG}$ & $f_{\rm enr}$\\\hline
Full & 8.1 & 42 & 0.016 & 1.3 \\
Half&  3.5 & 38 & 0.034 & 2.2 \\
$\gtrsim 90\%$ &  0.85 & 32 & 0.12 & 6.4 \\\hline

SIM2:C1 & $\rm Time=1620$ yr & $r\sim106$ AU & $\phi\sim 58^{\rm \circ}$ & $\rho_{\rm max}=1.7\times 10^{-11}$ g cm$^{-2}$\\\hline
Density Threshold &  Gas ($M_J$) & Rocks ($M_{\oplus}$) & $f_{\rm RtoG}$ & $f_{\rm enr}$\\\hline
Full & 8.5 & 42 & 0.015 & 1.3 \\
Half&  3.5 & 42 & 0.037 & 2.4 \\
$\gtrsim 90\%$ &  1.2 & 38 & 0.10 & 5.5  \\\hline\hline

SIM2km:C1 & $\rm Time=1430$ yr & $r\sim106$ AU & $\phi\sim 346^{\rm \circ}$ & $\rho_{\rm max}=1.3\times 10^{-11}$ g cm$^{-2}$\\\hline
Density Threshold &  Gas ($M_J$) & Rocks ($M_{\oplus}$) & $f_{\rm RtoG}$ & $f_{\rm enr}$\\\hline
Full & 7.1 & 7.2 & 0.0032 & 0.66 \\
Half&  3.7 & 2.6 & 0.0022 & 0.61  \\
$\gtrsim 90\%$ &  1.3 & 0.67 & 0.0016 & 0.58  \\\hline\hline

\end{tabular}
\end{table}

\begin{table}
\caption{Same as Table 1, but for clumps in SIM1mu.  Compare positions with Figure 6.}
\begin{tabular}{ l l l l  l }\hline

SIM1mu:C1 & $\rm Time=1100$ yr & $r\sim67$ AU & $\phi\sim 330^{\rm \circ}$ & $\rho_{\rm max}=2.3\times 10^{-11}$ g cm$^{-2}$\\\hline
Density Threshold &  Gas ($M_J$) & Rocks ($M_{\oplus}$) & $f_{\rm RtoG}$ & $f_{\rm enr}$\\\hline
Full & 8.2 & 55 & 0.021 & 1.5 \\
Half&  3.1 & 53 & 0.052 & 3.1 \\
$\gtrsim 90\%$ &  0.44 & 38 & 0.27 & 14 \\\hline\hline


SIM1mu:C2 & $\rm Time=1260$ yr & $r\sim87$ AU & $\phi\sim 51^{\rm \circ}$ & $\rho_{\rm max}=6.2\times 10^{-11}$ g cm$^{-2}$\\\hline
Density Threshold &  Gas ($M_J$) & Rocks ($M_{\oplus}$) & $f_{\rm RtoG}$ & $f_{\rm enr}$\\\hline
Full & 6.6 & 38 & 0.018 & 1.4 \\
Half&  2.6 & 27 & 0.033 & 2.1 \\
$\gtrsim 90\%$ &  0.66 & 12 & 0.059& 3.5 \\\hline

SIM1mu:C2 & $\rm Time=1410$ yr & $r\sim91$ AU & $\phi\sim 137^{\rm \circ}$ & $\rho_{\rm max}=3.2\times 10^{-11}$ g cm$^{-2}$\\\hline
Density Threshold &  Gas ($M_J$) & Rocks ($M_{\oplus}$) & $f_{\rm RtoG}$ & $f_{\rm enr}$\\\hline
Full & 11 & 70 & 0.020 & 1.5 \\
Half&  3.4 & 67 & 0.062 & 3.6 \\
$\gtrsim 90\%$ &  0.47 & 48 & 0.32 & 16 \\\hline

SIM1mu:C2 & $\rm Time=1570$ yr & $r\sim100$ AU & $\phi\sim 214^{\rm \circ}$ & $\rho_{\rm max}=4.6\times 10^{-11}$ g cm$^{-2}$\\\hline
Density Threshold &  Gas ($M_J$) & Rocks ($M_{\oplus}$) & $f_{\rm RtoG}$ & $f_{\rm enr}$\\\hline
Full & 15 & 83 & 0.017 & 1.4 \\
Half&  4.6 & 77 & 0.052 & 3.1 \\
$\gtrsim 90\%$ &  0.57 & 74 & 0.40& 21 \\\hline

SIM1mu:C2 & $\rm Time=1720$ yr & $r\sim104$ AU & $\phi\sim 276^{\rm \circ}$ & $\rho_{\rm max}=6.3\times 10^{-11}$ g cm$^{-2}$\\\hline
Density Threshold &  Gas ($M_J$) & Rocks ($M_{\oplus}$) & $f_{\rm RtoG}$ & $f_{\rm enr}$\\\hline
Full & 20 & 110& 0.018 & 1.4 \\
Half&  5.4 &86 & 0.050 & 3.0 \\
$\gtrsim 90\%$ &  0.53 & 80 & 0.47 & 24 \\

SIM1mu:C2 & $\rm Time=1870$ yr & $r\sim100$ AU & $\phi\sim 344^{\rm \circ}$ & $\rho_{\rm max}=1.2\times 10^{-10}$ g cm$^{-2}$\\\hline
Density Threshold &  Gas ($M_J$) & Rocks ($M_{\oplus}$) & $f_{\rm RtoG}$ & $f_{\rm enr}$\\\hline
Full & 32 & 170 & 0.017 & 1.3 \\
Half&  7.3 & 170 & 0.071 & 4.1 \\
$\gtrsim 90\%$ &  1.4 & 160 & 0.35& 18 \\\hline

\end{tabular}
\end{table}

\begin{table}
\caption{Same as Table 1, but for clump C3 in SIM1mu.  Compare positions with Figure 6.}
\begin{tabular}{ l l l l l }\hline

SIM1mu:C3 & $\rm Time=1410$ yr & $r\sim86$ AU & $\phi\sim 11^{\rm \circ}$ & $\rho_{\rm max}=4.2\times 10^{-11}$ g cm$^{-2}$\\\hline
Density Threshold &  Gas ($M_J$) & Rocks ($M_{\oplus}$) & $f_{\rm RtoG}$ & $f_{\rm enr}$\\\hline
Full & 10 & 54 & 0.017 & 1.4 \\
Half&  3.3 & 48 & 0.045 & 2.8 \\
$\gtrsim 90\%$ &  0.29 & 32 & 0.34 & 18 \\\hline

SIM1mu:C3 & $\rm Time=1570$ yr & $r\sim82$ AU & $\phi\sim 97^{\rm \circ}$ & $\rho_{\rm max}=4.7\times 10^{-11}$ g cm$^{-2}$\\\hline
Density Threshold &  Gas ($M_J$) & Rocks ($M_{\oplus}$) & $f_{\rm RtoG}$ & $f_{\rm enr}$\\\hline
Full & 11 & 61 & 0.017 & 1.4 \\
Half&  3.4 & 58 & 0.053 & 3.1 \\
$\gtrsim 90\%$ &  0.61 & 51 & 0.26 & 14 \\\hline

SIM1mu:C3 & $\rm Time=1720$ yr & $r\sim77$ AU & $\phi\sim 198^{\rm \circ}$ & $\rho_{\rm max}=3.0\times 10^{-11}$ g cm$^{-2}$\\\hline
Density Threshold &  Gas ($M_J$) & Rocks ($M_{\oplus}$) & $f_{\rm RtoG}$ & $f_{\rm enr}$\\\hline
Full & 10 & 64 & 0.020 & 1.5 \\
Half&  3.2 & 61 & 0.059 & 3.5 \\
$\gtrsim 90\%$ &  0.36 & 61 & 0.53 & 27 \\\hline

SIM1mu:C3 & $\rm Time=1870$ yr & $r\sim86$ AU & $\phi\sim 297^{\rm \circ}$ & $\rho_{\rm max}=4.3\times 10^{-11}$ g cm$^{-2}$\\\hline
Density Threshold &  Gas ($M_J$) & Rocks ($M_{\oplus}$) & $f_{\rm RtoG}$ & $f_{\rm enr}$\\\hline
Full & 11 & 64 & 0.018 & 1.4 \\
Half&  3.7 & 64 & 0.054 & 3.2 \\
$\gtrsim 90\%$ &  0.64 & 66 & 0.33 & 17 \\\hline

\end{tabular}
\end{table}

\end{document}